\documentclass[aps,pre,twocolumn,showpacs,10pt,superscriptaddress]{revtex4-1}

\usepackage{graphicx}
\usepackage{dcolumn}
\usepackage{bm}
\usepackage{amsmath}
\usepackage{xcolor}
\usepackage{amssymb}
\usepackage{float}
\usepackage[titletoc,title]{appendix}
\usepackage{soul}

\usepackage{ntheorem}
\newtheorem*{m_thm}{Main Theorem}

\newtheorem{thm}{Theorem}
\newtheorem{thm1}{Theorem}
\newtheorem*{thm2}{Theorem 2}

\newtheorem{conjecture_app}{Conjecture}

\newtheorem{conjecture}{Conjecture}
\newtheorem{fact}{Fact}
\newtheorem{lem}{Lemma}

\newcommand{\bra}[1]{\langle #1 |}
\newcommand{\ket}[1]{| #1 \rangle}
\newcommand{\shap}{\#\mathrm{P}}
\newcommand{\z}{\mathbf z}

\newcommand{\poly}[1]{\mathrm{poly}(#1)}

\begin{document}


\author{Jirawat Tangpanitanon}
\email{cqtjt@nus.edu.sg}
\affiliation{Centre for Quantum Technologies, National University of Singapore, 3 Science Drive 2, Singapore 117543}
\author{Supanut Thanasilp}
\affiliation{Centre for Quantum Technologies, National University of Singapore, 3 Science Drive 2, Singapore 117543}
\author{Marc-Antoine Lemonde}
\affiliation{Centre for Quantum Technologies, National University of Singapore, 3 Science Drive 2, Singapore 117543}
\author{Ninnat Dangniam}
\affiliation{
Department of Physics and Center for Field Theory and Particle Physics, Fudan University, Shanghai 200433, 
China}
\affiliation{State Key Laboratory of Surface Physics, Fudan University, Shanghai 200433, China}
\author{Dimitris G. Angelakis}
\email{dimitris.angelakis@qubit.org}
\affiliation{Centre for Quantum Technologies, National University of Singapore, 3 Science Drive 2, Singapore 117543}
\affiliation{School of Electrical and Computer Engineering, Technical University of Crete, Chania, Greece 73100}


\title{Quantum supremacy in driven quantum many-body systems }

\date{\today}

\begin{abstract}
A crucial milestone in the field of quantum simulation and computation is to demonstrate that a quantum device can perform a computation task that is classically intractable. 
A key question is to identify setups that can achieve such goal within current technologies. 
Here, we provide formal evidence that a large class of driven analog quantum systems constitute practical candidates for quantum supremacy. 
Our analysis is based on the Floquet eigenstate thermalization hypothesis, plausible assumptions and the absence of collapse of the polynomial hierarchy.
We give examples of driven disordered Ising chains and 1D driven Bose-Hubbard model. 
\end{abstract}

\maketitle

{\it Introduction--} Quantum computational supremacy is the ability of quantum devices to efficiently perform certain tasks that cannot be efficiently done on a classical computer \cite{2018_preskill_quantum,2017_Harrow_Nat}. Early proposals for realizing quantum supremacy include boson sampling \cite{2011_Aaronson_PTC,Aaronson2011,Ralph} and random quantum circuits~\cite{2011_shepherd, 2018_hartmut_natphy, Umesh}. In both cases, the computational hardness stems from the inability of a classical computer to efficiently sample the output probabilities of a complex quantum evolution. Experimental efforts towards achieving quantum supremacy include optical networks for boson sampling~\cite{2013_walmsley1_sci, 2013_Broome_Sci, 2013_walther_nat, 2013_fabio_natpho,PhysRevLett.123.250503} and superconducting circuits for random circuits~\cite{2018_neill_sci}. Signatures of quantum supremacy have been observed recently with 53 superconducting qubits \cite{2019_martinis_nat}. 

Analog quantum simulators are controllable quantum platforms specifically built to implement complex quantum many body models~\cite{2012_zoller_natphy, 2012_lewenstein_rpp,2014_dieter_epj,RevModPhys.86.153}. In these experiments, complex quantum dynamics have been implemented which cannot be reproduced with existing classical numerics and have shed light on important questions in quantum many-body physics~\cite{2016_Choi_Sci}. However, rigorous proof of quantum supremacy involving complexity theory in those systems 
are yet to be shown, with the few exceptions of the 2D quantum Ising~\cite{2018_eisert_prx,PhysRevLett.118.040502} and the 2D cluster-state models~\cite{2019_patron_arxiv}.

In this work, we provide evidence that when generic isolated periodically-driven quantum many-body systems thermalize, in the sense that any observables can be obtained from the microcanonical ensemble, sampling from their output distribution cannot be efficiently performed on a classical computer. These constitute a large class of quantum simulators that are currently available~\cite{2018_neill_sci, 2017_bloch_natphy, exp_cold_atoms_PhysRevX.9.041021,exp_cold_atoms_PhysRevX.10.011030, exp_trap_rev_Monroe2019ProgrammableQS,2017_Otterbach_ArXiV,2020arXiv200108226R}. 
Our analysis is based on the absence of collapse of the polynomial hierarchy and two plausible assumptions: the worst- to average-case hardness of the sampling task and the experimental realisability of random unitaries as predicted by the Floquet Eigenstate Thermalization Hypothesis (ETH).
We support our findings by examining specific examples of disordered quantum Ising chain driven by a global magnetic field and the one-dimensional Bose-Hubbard (BH) model with modulated hoppings. These models have been widely implemented experimentally~\cite{2018_neill_sci, 2017_bloch_natphy, exp_cold_atoms_PhysRevX.9.041021,exp_cold_atoms_PhysRevX.10.011030, exp_trap_rev_Monroe2019ProgrammableQS,2017_Otterbach_ArXiV,2020arXiv200108226R}, making our work of broad interest to the experimental community.


\textit{General framework--} Let us consider a generic periodically-driven quantum many-body system whose Hamiltonian is described by $\hat{H}(t)=\hat{H}_{0}+f(t)\hat{V}$. Here $\hat{H}_0$ is the undriven Hamiltonian, $\hat{V}$ is the driving Hamiltonian such that $\left[\hat{H}_0, \hat{V}\right]\neq 0$, and $f(t)$ is periodic with period $T$. We require that the time-averaged Hamiltonian $\hat{H}_{\rm ave}=\frac{1}{T}\int_0^T \hat{H}(t) dt$ describes an interacting many-body system~\cite{eisert}.

Let $\mathcal{Z}=\{|\bold{z}\rangle=\otimes_i^L |z_i\rangle\}$ be a complete basis of many-body Fock states, where $z_i=\{0,1,2,..,D_i-1\}$ denotes the basis state of a local quantum system of dimension $D_i$ and where $i \in [1,L]$. In what follows, we assume without loss of generality that $D_i=D$ for all $i$,
resulting in an Hilbert space of dimension $N=D^L$. The state after $M$ driving periods is $|\psi_M\rangle=\hat{U}_F^M|\bold{z}_0\rangle$, where $\hat{U}_F=\hat{\mathcal{T}}\exp\left(-i\int_0^T \hat{H}(t)dt\right)\equiv \exp\left (-i \hat{H}_F T\right)$ and $\hat{\mathcal{T}}$ is the time-ordering operator. We assume that the initial state $|\bold{z}_0\rangle$ is a product state. The effective time-independent Floquet Hamiltonian $\hat{H}_F$ fully describes the dynamics probed at stroboscopic times $t = nT$.  The probability of measuring the Fock state $|{\bf z}\rangle$ is then $p_M(\bold{z})=|\langle \bold{z} | \psi_M\rangle|^2$ with
\begin{equation}
\langle \bold{z} | \psi_M\rangle=\sum_{\bold{z}_1,...,\bold{z}_{M-1}\in \mathcal{Z}}   \prod_{m=0}^{M-1}\langle \bold{z}_{m+1} | \hat{U}_F| \bold{z}_{m}\rangle, 
\label{eq:path_integral}
\end{equation}
where the sum is performed over $M-1$ complete sets of basis states. More precisely, the set of basis states $\{ |{\bf z}_m \rangle\}$ is associated with the quantum evolution after $m$ driving cycles with ${\bf z}_0$ (${\bf z}_M=\bold{z}$) being the initial (readout) configuration. The expression in Eq. (\ref{eq:path_integral}) can be viewed as the Feynman's path integral where each trajectory is defined by a set of configurations $\{\bold{z}_{0}, \bold{z}_{1}, ...,\bold{z}_{M}\}$.

The ETH states that generic isolated many-body quantum systems thermalize by their own dynamics after a long enough time, regardless of their initial state. In that case, any generic observable is expected to evolve toward 
the canonical ensemble with a finite temperature~\cite{2016_Alessio_AiP}.
For driven quantum many-body systems, it has been shown that not only thermalization still occurs, but that for low-frequency driving, the associated temperature becomes infinite~\cite{2014_Rigol_PRX}. In this limit, the Floquet operator $\hat{U}_F$ shares the statistical properties of the Circular Orthogonal Ensemble (COE). This is an ensemble of matrices whose elements are independent normal complex random variables subjected to the orthogonality and the unitary constraints. This emergent randomness is the particular ingredient responsible for the hardness in calculating the output probability of  Eq. (\ref{eq:path_integral}), as there are exponentially many random Feynman trajectories that are equally important.
We emphasize that the external periodic drive is crucial to reach the required level of randomness~\cite{Kim2014, Lazarides2014}. A more detailed analysis of $\hat H_F$ shows that the presence of low-frequency driving allows to generate effective infinite-range multi-body interactions~\cite{2014_Rigol_PRX, Mori2016}. Therefore lifting the constraints imposed by the limited local few-body interactions generally encountered in physical systems~\footnote{Accurate descriptions of generic undriven thermalized systems using random matrix theory is in general only possible over small energy windows far from the energy-spectrum edges. If one analyses the entire energy spectrum, the local structure typically encountered in static Hamiltonians emerges and random matrix theory fails to capture it. This is not the case for driven thermalized systems which accurately applies to the entire $\hat U_F$ spectrum.}.

\textit{Quantum supremacy-- } To understand the computational task, let us first define some essential terms used in the complexity theory, namely \textit{approximating}, \textit{sampling}, \textit{multiplicative error} and \textit{additive error}. Let us imagine an analog quantum device built to mimic the quantum dynamics that would lead to $p_M(\bold{z})=|\langle \bold{z} | \psi_M\rangle|^2$. In practice, such device will encode an output probability $q(\z)$ that differs from $p_M(\bold{z})$ due to noise, decoherence and imperfect controls. Both probabilities are said to be multiplicatively close if
\begin{align}\label{eq:mult-sampling}
	|p_M(\bold{z})-q(\bold{z})|\leq  \alpha p_M(\bold{z})
\end{align} 
where $\alpha\geq 0$. The task of \textit{approximating $p_M(\bold{z})$ up to multiplicative error} is to calculate $q(\z)$ that satisfies the above equation for a given $\z$. However, such degree of precision is difficult to achieve experimentally as the allowed error is proportional to $p_M(\bold{z})$ which can be much smaller than unity. A more feasible task is to \textit{approximate $p_M(\bold{z})$ up to additive error}, defined as
\begin{align}\label{additive-sampling}
	\sum_{\bold{z}\in\mathcal{Z}}|p_M(\bold{z})-q(\bold{z})|\leq  \beta,
\end{align}
with $\beta>0$. Note that the additive error involves summing over all possible output strings $\z\in\mathcal{Z}$, while the multiplicative condition applies to each $\z$ individually.

The task of {\it approximating} $p_M(\z)$ even with additive error is still unrealistic as it requires a number of measurements that grows exponentially with the size of the system. What a quantum device can do is to sample strings from $q(\bold{z})$. Hence, we define the task of \textit{sampling from $p_M(\z)$ up to additive error} as generating strings from $q(\bold{z})$ while $q(\bold{z})$ is additively close to $p_M(\z)$. 
This task is our central focus to show quantum supremacy.
We emphasize that it is different from ``certifying quantum supremacy'' \cite{2020_jens} which consists of certifying if Eq.~(\ref{additive-sampling}) holds.

To show that the above sampling task cannot be done efficiently by a classical computer, we follow the standard argument which proceeds as follows. Let us suppose that there is a classical machine $\mathcal{C}$ able to \textit{sample from $p_M(\z)$ up to additive error} and that the distribution of $p_M(\bold{z})$ anti-concentrates, i.e. 
\begin{equation}
{\rm Pr}\left(p_M(\bold{z}) > \frac{\delta}{N}\right) \ge \gamma,
\label{eq:anticoncentration}
\end{equation} 
for some positive constants $\delta,\gamma >0$ for all $\z\in\mathcal{Z}$ \cite{Hangleiter2018anticoncentration}. Here, the distribution is obtained from a set of unitary matrices $\{\hat{U}_F\}$ that are realizable experimentally. The Stockmeyer theorem states that, with the help of a NP oracle, that machine $\mathcal{C}$ can also \textit{approximate $p_M(\bold{z})$ up to multiplicative error} for some outcomes $\z$~\cite{doi:10.1137/0214060}. We emphasize that the {\it sampling} task is converted to the {\it approximation} task in this step. If the latter is $\#$P-hard, then the existence of that machine $\mathcal{C}$ would imply the collapse of the polynomial hierarchy to the third level, which is strongly believed to be unlikely in computer science. Hence, assuming that the polynomial hierarchy does not collapse to the third level, we reach the conclusion that a classical machine $\mathcal{C}$ does not exist. 

The two fundamental conditions of the proof, that is the $\#$P-hardness of \textit{approximating $p_M(\bold{z})$ up to multiplicative error} and the anti-concentration of $p_M(\bold{z})$, can be more formally based on the two following theorems.

\begin{thm}\label{thm:sharp_P} Let $\mathcal{Y}$ be a set of output probabilities $\tilde{p}_M(\bold{z})=|\langle \bold{z}|\hat{U}^M_{\rm COE}|\bold{z}_0\rangle|^2$ obtained from all possible COE matrices $\{\hat{U}_{\rm COE}\}$ and all possible output strings $\z\in\mathcal{Z}$. Approximating $\tilde{p}_M(\bold{z})$ in $\mathcal{Y}$ up to multiplicative error is $\shap$ hard in the worst case.
\end{thm}

\begin{thm}\label{thm:anticoncentration} The distribution of $\tilde{p}_M(\bold{z})$ in $\mathcal{Y}$ anti-concentrates with $\delta =1$ and $\gamma=1/e$, where $e$ is the base of the natural logarithm.
\end{thm}
In theorem~\ref{thm:sharp_P}, we introduced the key notion of \textit{worst-case} hardness of the entire set of COE matrices $\{\hat{U}_{\rm COE}\}$.  
This corresponds to the scenario where at least one instance $\tilde{p}_M(\bold{z})$, i.e.~a single unitary $\hat U \in \{\hat{U}_{\rm COE}\}$ and a single configuration $\z\in\mathcal{Z}$, is hard to approximate with multiplicative error.
However, that one instance may be impractical to produce experimentally as the full set of COE matrices $\{\hat{U}_{\rm COE}\}$ ($\tilde{p}_M(\z)$) might not coincide with the experimentally accessible set $\{\hat{U}_F\}$ ($p_M(\z)$). 
That is due to  the fact that even though Floquet ETH strongly suggests that $\hat{U}^M_F$ is an instance uniformly drawn from the $\{\hat{U}^M_{\rm COE}\}$, not all $\hat{U}^M_{\rm COE}$ matrices will be realizable by $\{\hat{U}^M_F\}$.
More desirable is the \textit{average-case} hardness where most instances are hard.
Consequently, to ensure that the hard instance in $\mathcal{Y}$ can be found within $\{\hat{U}_F\}$ and thus prove quantum supremacy for realizable driven analog quantum systems, we further assume the following two conjectures.

\begin{conjecture}[Average-case hardness]\label{conj:average-hard} For any $1/2e$ fraction of $\mathcal{Y}$, approximating $\tilde{p}_M(\z)$ up to multiplicative error with $\alpha= 1/4+o(1)$ is as hard as the hardest instance. Here $o(\cdot)$ is the little-o notation.
\end{conjecture}

\begin{conjecture}[Computational Floquet ETH]\label{conj:ETH}
The experimentally accessible set $\{\hat{U}_F^M\}$ is approximately Haar random over the ensemble $\{\hat{U}_{\rm COE}^M\}$ in the sense that (1) the distribution of $p_M(\z)$ over $\{\hat{U}_F^M\}$ is equivalent to that of $\tilde{p}_M(\z)$ over $\{\hat{U}_{\rm COE}^M\}$ and (2) the average instances are as hard in both ensembles.
\end{conjecture}

Informally, conjecture \ref{conj:average-hard} assumes the worst-to-average case reduction in $\mathcal{Y}$ which is common in most quantum supremacy proposals~\footnote{Even though there are recent breakthroughs worst-to-average case reduction for hardness of computing output probabilities of quantum circuits. So far none of the approaches matches the realistic error requirement to rule out a classical sampler from the Stockmeyer argument \cite{bouland2019complexity, movassagh_cayley_2019}}. 
Conjecture~\ref{conj:ETH} connects the mathematically constructed COE to experimentally accessible driven analog quantum systems by stating that the ensemble $\{\hat{U}_{F}^M\}$ is statistically equivalent to a set of instances uniformly drawn from $\{\hat{U}_{\rm COE}^M\}$. 
This conjecture is supported by the observation that isolated systems evolving under generic $\hat{U}_{F}^M$ thermalize to infinite temperature resulting in fully random final quantum states~\cite{Kim2014, Lazarides2014, 2014_Rigol_PRX} in experimentally relevant timescales~\cite{2017_bloch_natphy, Mori2016}.
Compared to existing quantum supremacy proposals, the reliance of the main theorem (see below) on conjecture~\ref{conj:ETH} is not standard and may be seen as undesirable. But from our perspective, this conjecture makes a connection between computational complexity and the experimentally tested Floquet ETH that is applicable to a broad class of generic periodically-driven quantum systems as implemented in a variety of analog quantum simulators. Proving or disproving conjecture~\ref{conj:ETH}, either directly or indirectly by refutation of the main theorem while conjecture~\ref{conj:average-hard} holds true, is by itself of fundamental interest in physics.

The fraction used in conjecture~\ref{conj:average-hard} is chosen so that the approximate Haar random measure ensures  that some hard instances in $\mathcal{Y}$ can be realized with $\{\hat{U}_F^M\}$.
In combination of theorems~\ref{thm:sharp_P} and \ref{thm:anticoncentration}, the two conjectures finally allow us to state the main theorem of this work. 
\begin{m_thm}\label{thm:main}
	Assuming conjectures \ref{conj:average-hard} and \ref{conj:ETH}, the ability to classically sample from $p_M(\z)$ up to an additive error $\beta=1/8e$ for all unitary matrices in $\{\hat{U}_F\}$ implies the collapse of the polynomial hierarchy to the third level.
\end{m_thm}
In what follows, we address in detail the proofs of theorems~\ref{thm:sharp_P} and \ref{thm:anticoncentration} while the detailed application of the standard Stockmeyer argument to prove the main theorem is provided in Appendix~\ref{app:average-hardness} 


\textit{\#P hardness of simulating COE quantum dynamics-- } To prove theorem~\ref{thm:sharp_P}, we first notice that the COE is an ensemble of all orthogonal unitary matrices. This includes the well-known instantaneous quantum polynomial (IQP) circuits $\hat{U}_{\rm IQP}=\hat{\mathcal{H}}\hat{\mathcal{Z}}\hat{\mathcal{H}}$, where $\hat{\mathcal{H}}$ consists of Hadamard gates and $\hat{\mathcal{Z}}$ is an arbitrary (possibly non-local) diagonal gate on the computational basis, both acting on all qubits~\cite{2011_shepherd}. The IQP circuits constitute one of the early proposals of quantum supremacy. Multiplicative approximation of their output probabilities are known to be $\#$P-hard in the worst case~\cite[Theorem 1.4]{goldberg_guo_2017}.  
Since $\hat{U}^M_{\rm IQP}=\hat{\mathcal{H}}\hat{\mathcal{Z}}^M\hat{\mathcal{H}}$ still adopt the general form of the IQP circuits, we conclude that there exists at least one instance in $\mathcal{Y}$ that is $\#$P-hard for multiplicative approximation. 

To see how the hardness could emerge for a typical instance in $\mathcal{Y}$ (conjecture \ref{conj:average-hard}), one can in principle map the path integral in Eq.~(\ref{eq:path_integral}) to the partition function of a classical Ising model with random complex fields. The latter is widely conjectured to be \#P-hard on average for multiplicative approximation~\cite{2018_eisert_prx,bremner_average_2016}. 
In this context, the key is to note that a COE unitary evolution can be written as $\hat{U}_{\rm COE}=\hat{U}^T_{\rm CUE}\hat{U}_{\rm CUE}$, where $\hat{U}_{\rm CUE}$ is a random matrix drawn from the Circular Unitary Ensemble (CUE), i.e.~the ensemble of Haar-random matrices \cite{2010_haake}.
Furthermore, $\hat{U}_{\rm CUE}$ can be decomposed into a set of universal quantum gates which can be mapped onto a complex Ising model.
This mapping procedure has already been described in ref.~\cite{2018_hartmut_natphy} to support the conjecture of the worst-to-average case in the context of random quantum circuits. 
A detailed and intuitive description of this protocol is presented in Appendix~\ref{app:mapping}.

\begin{figure}
\includegraphics[width=1.0\columnwidth]{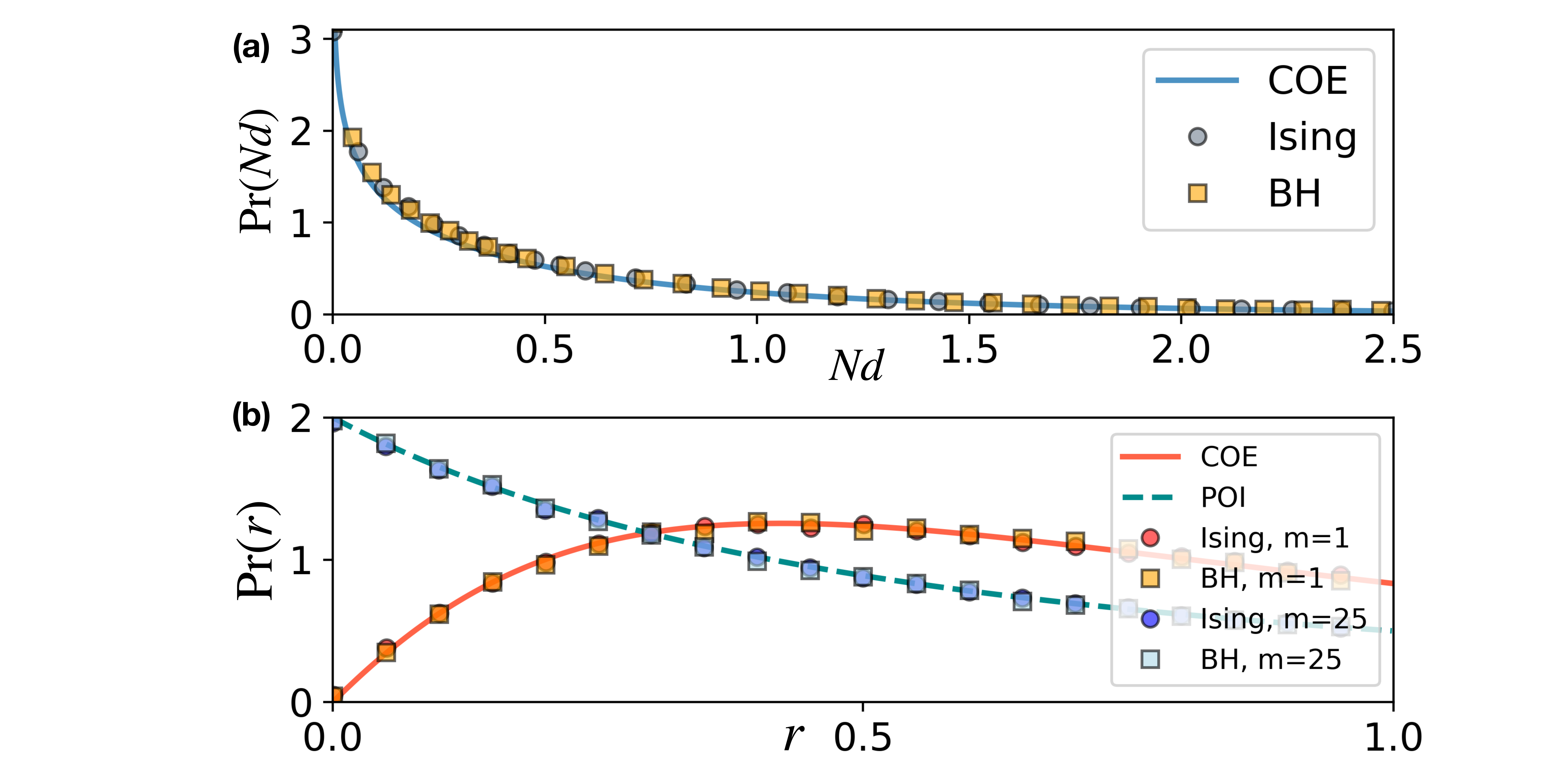}
\caption{(a) The eigenstate distribution $d_{\epsilon}(\bold{z})$ for the Ising and the BH models. The blue line is the Bessel function of the second kind predicted by COE. (b) The statistics of level spacings obtained from the Ising and the BH chain at $M=1,25$. The blue dashed and the solid lines are the POI and the COE distributions, respectively.  Ising and BH parameters: $L = 10$ (with half-filling for the BH model), $W=1J, F=2.5J, \omega=8J$, and $500$ disorder realizations.}
\label{fig2}
\end{figure}

\textit{Anti-concentration of COE dynamics.- } To prove the second and necessary ingredient of the proof, i.e.~theorem~\ref{thm:anticoncentration}, we write
\begin{equation}
\langle \bold{z} |\hat{U}^M_{\rm COE}|\z_0\rangle = \sum_{\epsilon=0}^{N-1}d_{\epsilon}(\bold{z}) e^{i\phi_{M,\epsilon}}, 	
\end{equation}
where $d_{\epsilon}(\bold{z})=\langle \bold{z} | E_\epsilon \rangle\langle E_\epsilon |\bold{z}_0\rangle$, $\phi_{M,\epsilon}= ME_{\epsilon}T \text{ mod } 2\pi$, $|E_{\epsilon}\rangle$ is an eigenstate of $\hat{H}_F$ with eigenenergy $E_\epsilon$. For COE operators, $d_\epsilon({\bf z})$ are real \cite{2010_haake} and their distribution, denoted as Pr$(d)$, is given by the Bessel function of the second kind (see Fig.~\ref{fig2}(a) and 
Appendix~\ref{app:anti-concentration} for a detailed derivation). Consequently, the values of $d_{\epsilon}(\bold{z})$ for different $\epsilon$ and $\bold{z}$ do not concentrate on a particular value. 

Now let us consider the statistics of the phases $\{\phi_{M,\epsilon}\}$. We define the level spacing as $r_{\epsilon}={\rm min}(\delta_{\epsilon+1},\delta_{\epsilon})/{\rm max}(\delta_{\epsilon+1},\delta_{\epsilon})$ with $\delta_{\epsilon}=\phi_{\epsilon+1}-\phi_{\epsilon}>0$. For a single driving cycle $M=1$, the phases $\{\phi_{1,\epsilon}\}$ for COE are known to exhibit phase repulsion, i.e. the phases are correlated \cite{2014_Rigol_PRX}. The COE distribution ${\rm Pr}_{\rm COE}(r_{\epsilon})$ is depicted in Fig.~\ref{fig2}(b), where ${\rm Pr}_{\rm COE}(0) = 0$ explicitly indicates the phase repulsion.
For multiple driving cycles $M \gg 2\pi/ E_{\epsilon}T$, the correlations are erased due to energy folding, i.e.~the effect of the modulo $2\pi$. This results in the Poisson (POI) distribution of the level spacing, 
${\rm Pr}_{\rm POI}(r_{\epsilon}) = 2/(1+r_{\epsilon}^2)$, with the peak at $r=0$, see Fig.~\ref{fig2}(b).

The Bessel function distribution of $d_{\epsilon}(\bold{z})$ and the POI distribution of $\phi_{M,\epsilon}$ ensure that the output distribution ${\rm Pr}(p)$ is not concentrated. Specially, ${\rm Pr}(p)$ follows the so-called Porter-Thomas distribution ${\rm Pr}_{\rm PT}(p) = Ne^{-Np}$, which suggests that the system explores uniformly (approximately Haar-random) the Hilbert space~\cite{2018_neill_sci, 2018_eisert_prx}. This satisfies the anti-concentration condition since ${\rm Pr}_{\rm PT}\left(p > \frac{1}{N}\right) = \int_{Np=1}^{\infty} d(Np) e^{-Np} = 1/e$ \cite{2018_hartmut_natphy}. To see the emergence of the Porter-Thomas distribution, we write $\langle \bold{z} |\psi_M\rangle = a_{\bold{z}}+ib_{\bold{z}}$, where $a_{\bold{z}}=\sum_{\epsilon}d_{\epsilon}(\bold{z})\cos \phi_{M,\epsilon}$ and $b_{\bold{z}}=\sum_{\epsilon}d_{\epsilon}(\bold{z})\sin \phi_{M,\epsilon}$. Due to the Poisson distribution in the long time limit, the phases $\{\phi_{M,\epsilon}\}$ can be thought of as independent variables randomly and uniformly distributed in the range $\left[0,2\pi\right)$. Using the product distribution formula and the central limit theorem, one can show that the distributions of $a_{\bold{z}}$ and $b_{\bold{z}}$ are normal distributions with zero mean and variance $1/2N$. Since $\tilde p_M(\bold{z})=a^2_{\bold{z}}+b^2_{\bold{z}}$, the Porter-Thomas distribution of $\tilde p_M(\bold{z})$ can be derived using the fact that the square sum of two Gaussian variables follows the $\chi$-squared distribution with second degree of freedom \cite{2002_simon}.
A detailed derivation is presented in Appendix~\ref{app:anti-concentration}. 


\begin{figure}
\includegraphics[width=0.95\columnwidth]{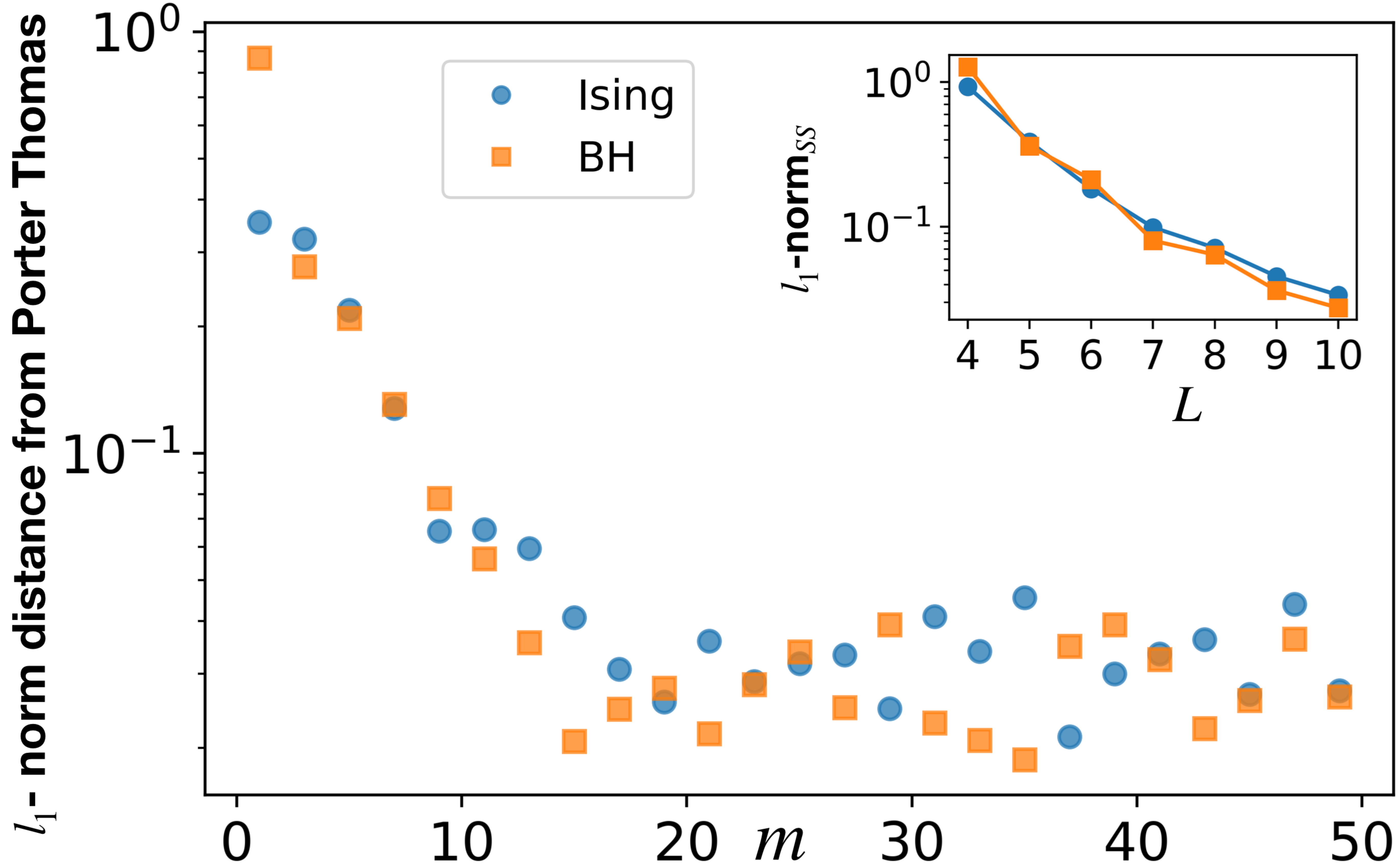}
\caption{The $l_1$-norm distance between the output distribution from different quantum systems and the Porter-Thomas distribution at different $m$. The results from the Ising chain and the BH chain are labeled as circles and squares, respectively. Ising and BH parameters: $L = 10$ (with half-filling for the BH model), $W=1J, F=2.5J, \omega=8J$, and $500$ disorder realizations. The inset shows the plot of $l_1$-norm distance in the long time limit as a function of $L$.}
\label{fig3}
\end{figure}

\textit{Example of driven many-body systems.-} We give two specific examples of driven systems that display statistical properties consistent with the COE and hence partially support conjecture~\ref{conj:ETH}. For both cases, the modulation is $f(t) = \frac{1}{2}(1-\cos(\omega t))$, where $\omega=2\pi/T$ and initial states are randomized product states. 

(i) {1D Ising chain:} We consider an Ising chain described by the Hamiltonian $\hat{H}_0^{\rm ISING}=\sum_{l=0}^{L-1} \mu_l \hat{Z}_l + J\sum_{l=0}^{L-2}\hat{Z}_l\hat{Z}_{l+1}$, where $\mu_l\in\{0,W\}$ is a local disorder, $W$ is the disorder strength, $\hat{Z}_l$ is the Pauli spin operator acting on site $l$, and $J$ is the interaction strength. The drive is a global magnetic field $\hat{V}^{\rm ISING}=F\sum_{l=0}^{L-1}\hat{X}_l$, where $F$ is the driving amplitude. Similar models have been implemented in various quantum platforms, including trapped ions \cite{exp_trap_rev_Monroe2019ProgrammableQS} and superconducting circuits \cite{2017_Otterbach_ArXiV}. 

(ii) {1D Bose-Hubbard model:} We consider the BH model described by the Hamiltonian $\hat{H}^{\rm BH}_0=\sum_{l=0}^{L-1}(\mu_l \hat{a}^\dagger_l \hat{a}_l+\frac{U}{2}\hat{a}^\dagger_l\hat{a}^\dagger_l\hat{a}_l\hat{a}_l)$, where $\hat{a}_l$ ($\hat{a}^\dag_l$) is a bosonic annihilation (creation) operator at site $l$, $U$ is the on-site interaction, and $\mu_l$ is the local disorder as defined above. The drive modulates the hopping amplitudes $\hat{V}^{\rm BH}=- F \sum_{l=0}^{L-2}(\hat{a}^\dagger_l\hat{a}_{l+1}+ {\rm H.c.})$. Similar models have been implemented in superconducting circuits~\cite{2018_neill_sci} and cold atoms \cite{2017_bloch_natphy, 2020arXiv200108226R,exp_cold_atoms_PhysRevX.10.011030}.

The distribution of $d_{\epsilon}(\bold{z})$ from both models are depicted in Fig.~\ref{fig2}(a), showing an agreement with the Bessel function as predicted by COE. The level statistics at $M=1$ and $M=25$ are depicted in Fig.~\ref{fig2}(b), showing an agreement with the COE and the POI distribution, respectively. The driving frequency and the disorder strength are tuned to ensure the observation of the thermalized phase and prevent many-body localization \cite{p2, 2014_Rigol_PRX}.

Fig. \ref{fig3} shows the $l_1$-norm distance between ${\rm Pr}(p)$ and the Porter-Thomas distribution at different $m$ for the Ising and the BH models. It can be seen that, in all cases, the system reaches the Porter-Thomas distribution after multiple driving cycles. The $l_1$-norm distance in the long-time limit is decaying towards zero as the size of the system increases. Therefore, the anti-concentration condition is satisfied.
In absence of the drive, a similar analysis can be performed for the infinite-time unitary evolution corresponding to generic instances of the undriven thermalized phase in both models. In this case, $d_{\epsilon}(\bold{z})$ does not follow the Bessel function of the second kind and the output distribution never reaches the Porter-Thomas distribution~(see Appendix~\ref{app:undriven_physical} for numerical simulation of the undriven Ising and Bose-Hubbard models). This is consequence of the energy conservation and the structure imposed by the local interactions, highlighting the key role played by the drive.


{\it Conclusions and outlook--} Analog quantum simulators realizing quantum many-body systems have generated quantum dynamics beyond the reach of existing classical numerical methods for some time. However, such dynamics has not been theoretically proven to be hard to compute by a classical computer. We have shown here that in the particular case of driven many-body systems, when they thermalize, sampling from their output distribution cannot be efficiently performed on a classical computer. Using complexity theory arguments, we provide strong analytical evidence of the computational hardness stemming from the COE statistics, and provide numerical results showing that COE dynamics can be obtained from driven quantum Ising and BH models for realistic parameters. 

Our results greatly widen the possibilities to realise quantum supremacy with existing experimental platforms and provide the theoretical foundations needed to demonstrate quantum supremacy in analog quantum simulators. In the future, it would be interesting to extend our results to a broader class of quantum many-body systems such as those with gauge fields, frustrated spin systems, and undriven systems. For example, in Ref.~\cite{2016_Choi_Sci}, cold atoms in optical lattices have been used to compute the undriven quantum many-body localization transition in two dimensions, which has so far eluded state-of-the- art classical numerical techniques~\cite{doi:10.1080/14789940801912366}.
  



{\it Acknowledgement--}
This research is supported by the National Research Foundation, Prime Minister's Office, Singapore and the Ministry of Education, Singapore under the Research Centres of Excellence programme.
It was also partially funded by Polisimulator project co-financed by Greece and the EU Regional Development Fund. Ninnat Dangniam is supported by the National Natural Science Foundation
of China (Grant No. 11875110).

\phantom{\cite{bouland2019complexity, movassagh_cayley_2019}} 
\bibliography{ref}

\begin{thebibliography}{10}

\bibitem{2018_preskill_quantum}
John Preskill.
\newblock Quantum {C}omputing in the {NISQ} era and beyond.
\newblock {\em {Quantum}}, 2:79, August 2018.

\bibitem{2017_Harrow_Nat}
A.~W. Harrow and A.~Montanaro.
\newblock Quantum computational supremacy.
\newblock {\em Nature}, 549:203, 09 2017.

\bibitem{2011_Aaronson_PTC}
S.~Aaronson and A.~Arkhipov.
\newblock The computational complexity of linear optics.
\newblock {\em Proceedings of the 43rd annual ACM Symposium on Theory of
  Computing, (STOC '11)}, pages 333--342, 2011.

\bibitem{Aaronson2011}
Aaronson Scott.
\newblock A linear-optical proof that the permanent is np-hard.
\newblock {\em Proceedings of the Royal Society A: Mathematical, Physical and
  Engineering Sciences}, 467:3393, 2011.

\bibitem{Ralph}
A.~P. Lund, Michael~J. Bremner, and T.~C. Ralph.
\newblock Quantum sampling problems, bosonsampling and quantum supremacy.
\newblock {\em npj Quantum Information}, 3(1):15, 2017.

\bibitem{2011_shepherd}
Michael~J. Bremner, Richard Jozsa, and Dan~J. Shepherd.
\newblock Classical simulation of commuting quantum computations implies
  collapse of the polynomial hierarchy.
\newblock {\em Proceedings of the Royal Society A: Mathematical, Physical and
  Engineering Sciences}, 467(2126):459--472, 2011.

\bibitem{2018_hartmut_natphy}
S.~Boixo, S.~V. Isakov, V.~N. Smelyanskiy, R.~Babbush, N.~Ding, Z.~Jiang, M.~J.
  Bremner, J.~M. Martinis, and H.~Neven.
\newblock Characterizing quantum supremacy in near-term devices.
\newblock {\em Nature Physics}, 14(6):595--600, 2018.

\bibitem{Umesh}
Adam Bouland, Bill Fefferman, Chinmay Nirkhe, and Umesh Vazirani.
\newblock On the complexity and verification of quantum random circuit
  sampling.
\newblock {\em Nature Physics}, 15(2):159--163, 2019.

\bibitem{2013_walmsley1_sci}
J.~B. Spring, B.~J. Metcalf, P.~C. Humphreys, W.~S. Kolthammer, X.-M. Jin,
  M.~Barbieri, A.~Datta, N.~Thomas-Peter, N.~K. Langford, D.~Kundys, J.~C.
  Gates, B.~J. Smith, P.~G.~R. Smith, and I.~A. Walmsley.
\newblock Boson sampling on a photonic chip.
\newblock {\em Science}, 339(6121):798--801, 2013.

\bibitem{2013_Broome_Sci}
M.~A. Broome, A.~Fedrizzi, S.~Rahimi-Keshari, J.~Dove, S.~Aaronson, T.~C.
  Ralph, and A.~G. White.
\newblock Photonic boson sampling in a tunable circuit.
\newblock {\em Science}, 339(6121):794--798, 2013.

\bibitem{2013_walther_nat}
M.~Tillmann, B.~Daki{\'c}, R.~Heilmann, S.~Nolte, A.~Szameit, and P.~Walther.
\newblock Experimental boson sampling.
\newblock {\em Nature Photonics}, 7:540, 05 2013.

\bibitem{2013_fabio_natpho}
A.~Crespi, R.~Osellame, R.~Ramponi, D.~J. Brod, E.~F. Galv{\~a}o, N.~Spagnolo,
  C.~Vitelli, E.~Maiorino, P.~Mataloni, and F.~Sciarrino.
\newblock Integrated multimode interferometers with arbitrary designs for
  photonic boson sampling.
\newblock {\em Nature Photonics}, 7:545, 05 2013.

\bibitem{PhysRevLett.123.250503}
Hui Wang, Jian Qin, Xing Ding, Ming-Cheng Chen, Si~Chen, Xiang You, Yu-Ming He,
  Xiao Jiang, L.~You, Z.~Wang, C.~Schneider, Jelmer~J. Renema, Sven H\"ofling,
  Chao-Yang Lu, and Jian-Wei Pan.
\newblock Boson sampling with 20 input photons and a 60-mode interferometer in
  a $1{0}^{14}$-dimensional hilbert space.
\newblock {\em Phys. Rev. Lett.}, 123:250503, Dec 2019.

\bibitem{2018_neill_sci}
C.~Neill, P.~Roushan, K.~Kechedzhi, S.~Boixo, S.~V. Isakov, V.~Smelyanskiy,
  A.~Megrant, B.~Chiaro, A.~Dunsworth, K.~Arya, R.~Barends, B.~Burkett,
  Y.~Chen, Z.~Chen, A.~Fowler, B.~Foxen, M.~Giustina, R.~Graff, E.~Jeffrey,
  T.~Huang, J.~Kelly, P.~Klimov, E.~Lucero, J.~Mutus, M.~Neeley, C.~Quintana,
  D.~Sank, A.~Vainsencher, J.~Wenner, T.~C. White, H.~Neven, and J.~M.
  Martinis.
\newblock A blueprint for demonstrating quantum supremacy with superconducting
  qubits.
\newblock {\em Science}, 360(6385):195--199, 2018.

\bibitem{2019_martinis_nat}
Frank Arute, Kunal Arya, Ryan Babbush, Dave Bacon, Joseph~C. Bardin, Rami
  Barends, Rupak Biswas, Sergio Boixo, Fernando G. S.~L. Brandao, David~A.
  Buell, Brian Burkett, Yu~Chen, Zijun Chen, Ben Chiaro, Roberto Collins,
  William Courtney, Andrew Dunsworth, Edward Farhi, Brooks Foxen, Austin
  Fowler, Craig Gidney, Marissa Giustina, Rob Graff, Keith Guerin, Steve
  Habegger, Matthew~P. Harrigan, Michael~J. Hartmann, Alan Ho, Markus Hoffmann,
  Trent Huang, Travis~S. Humble, Sergei~V. Isakov, Evan Jeffrey, Zhang Jiang,
  Dvir Kafri, Kostyantyn Kechedzhi, Julian Kelly, Paul~V. Klimov, Sergey Knysh,
  Alexander Korotkov, Fedor Kostritsa, David Landhuis, Mike Lindmark, Erik
  Lucero, Dmitry Lyakh, Salvatore Mandr{\`a}, Jarrod~R. McClean, Matthew
  McEwen, Anthony Megrant, Xiao Mi, Kristel Michielsen, Masoud Mohseni, Josh
  Mutus, Ofer Naaman, Matthew Neeley, Charles Neill, Murphy~Yuezhen Niu, Eric
  Ostby, Andre Petukhov, John~C. Platt, Chris Quintana, Eleanor~G. Rieffel,
  Pedram Roushan, Nicholas~C. Rubin, Daniel Sank, Kevin~J. Satzinger, Vadim
  Smelyanskiy, Kevin~J. Sung, Matthew~D. Trevithick, Amit Vainsencher, Benjamin
  Villalonga, Theodore White, Z.~Jamie Yao, Ping Yeh, Adam Zalcman, Hartmut
  Neven, and John~M. Martinis.
\newblock Quantum supremacy using a programmable superconducting processor.
\newblock {\em Nature}, 574(7779):505--510, 2019.

\bibitem{2012_zoller_natphy}
J.~I. Cirac and P.~Zoller.
\newblock Goals and opportunities in quantum simulation.
\newblock {\em Nat Phys}, 8:264, 04 2012.

\bibitem{2012_lewenstein_rpp}
P.~Hauke, F.~M. Cucchietti, L.~Tagliacozzo, I.~Deutsch, and M.~Lewenstein.
\newblock Can one trust quantum simulators?
\newblock {\em Reports on Progress in Physics}, 75(8):082401, 2012.

\bibitem{2014_dieter_epj}
T.~H. Johnson, S.~R. Clark, and D.~Jaksch.
\newblock What is a quantum simulator?
\newblock {\em EPJ Quantum Technology}, 1(1):10, Jul 2014.

\bibitem{RevModPhys.86.153}
I.~M. Georgescu, S.~Ashhab, and Franco Nori.
\newblock Quantum simulation.
\newblock {\em Rev. Mod. Phys.}, 86:153--185, Mar 2014.

\bibitem{2016_Choi_Sci}
J.-Y. Choi, S.~Hild, J.~Zeiher, P.~Schau{\ss}, A.~Rubio-Abadal, T.~Yefsah,
  V.~Khemani, D.~A. Huse, I.~Bloch, and C.~Gross.
\newblock Exploring the many-body localization transition in two dimensions.
\newblock {\em Science}, 352(6293):1547--1552, 2016.

\bibitem{2018_eisert_prx}
Juan Bermejo-Vega, Dominik Hangleiter, Martin Schwarz, Robert Raussendorf, and
  Jens Eisert.
\newblock Architectures for quantum simulation showing a quantum speedup.
\newblock {\em Phys. Rev. X}, 8:021010, Apr 2018.

\bibitem{PhysRevLett.118.040502}
Xun Gao, Sheng-Tao Wang, and L.-M. Duan.
\newblock Quantum supremacy for simulating a translation-invariant ising spin
  model.
\newblock {\em Phys. Rev. Lett.}, 118:040502, Jan 2017.

\bibitem{2019_patron_arxiv}
Leonardo {Novo}, Juani {Bermejo-Vega}, and Ra{\'u}l {Garc{\'\i}a-Patr{\'o}n}.
\newblock {Quantum advantage from energy measurements of many-body quantum
  systems}.
\newblock {\em arXiv e-prints}, page arXiv:1912.06608, Dec 2019.

\bibitem{2017_bloch_natphy}
Pranjal Bordia, Henrik L{\"u}schen, Ulrich Schneider, Michael Knap, and
  Immanuel Bloch.
\newblock Periodically driving a many-body localized quantum system.
\newblock {\em Nature Physics}, 13(5):460--464, 2017.

\bibitem{exp_cold_atoms_PhysRevX.9.041021}
K.~Singh, C.~J. Fujiwara, Z.~A. Geiger, E.~Q. Simmons, M.~Lipatov, A.~Cao,
  P.~Dotti, S.~V. Rajagopal, R.~Senaratne, T.~Shimasaki, M.~Heyl, A.~Eckardt,
  and D.~M. Weld.
\newblock Quantifying and controlling prethermal nonergodicity in interacting
  floquet matter.
\newblock {\em Phys. Rev. X}, 9:041021, Oct 2019.

\bibitem{exp_cold_atoms_PhysRevX.10.011030}
K.~Wintersperger, M.~Bukov, J.~N\"ager, S.~Lellouch, E.~Demler, U.~Schneider,
  I.~Bloch, N.~Goldman, and M.~Aidelsburger.
\newblock Parametric instabilities of interacting bosons in periodically driven
  1d optical lattices.
\newblock {\em Phys. Rev. X}, 10:011030, Feb 2020.

\bibitem{exp_trap_rev_Monroe2019ProgrammableQS}
C.~Monroe, W.~C. Campbell, L.~Duan, Z-X Gong, A.~V. Gorshkov, P.~Hess,
  R.~Islam, K.~Kim, N.~Linke, G.~Pagano, P.~Richerme, C.~Senko, and N.~Yao.
\newblock Programmable quantum simulations of spin systems with trapped ions.
\newblock {\em arXiv preprint, arXiv:1912.07845}, 2019.

\bibitem{2017_Otterbach_ArXiV}
J.~S. {Otterbach}, R.~{Manenti}, N.~{Alidoust}, A.~{Bestwick}, M.~{Block},
  B.~{Bloom}, S.~{Caldwell}, N.~{Didier}, E.~{Schuyler Fried}, S.~{Hong},
  P.~{Karalekas}, C.~B. {Osborn}, A.~{Papageorge}, E.~C. {Peterson},
  G.~{Prawiroatmodjo}, N.~{Rubin}, Colm~A. {Ryan}, D.~{Scarabelli},
  M.~{Scheer}, E.~A. {Sete}, P.~{Sivarajah}, Robert~S. {Smith}, A.~{Staley},
  N.~{Tezak}, W.~J. {Zeng}, A.~{Hudson}, Blake~R. {Johnson}, M.~{Reagor}, M.~P.
  {da Silva}, and C.~{Rigetti}.
\newblock {Unsupervised Machine Learning on a Hybrid Quantum Computer}.

\bibitem{2020arXiv200108226R}
Antonio {Rubio-Abadal}, Matteo {Ippoliti}, Simon {Hollerith}, David {Wei}, Jun
  {Rui}, S.~L. {Sondhi}, Vedika {Khemani}, Christian {Gross}, and Immanuel
  {Bloch}.
\newblock {Floquet prethermalization in a Bose-Hubbard system}.
\newblock {\em arXiv e-prints}, page arXiv:2001.08226, Jan 2020.

\bibitem{eisert}
J.~Eisert, Friesdorf M., and C.~Gogolin.
\newblock Quantum many-body systems out of equilibrium.
\newblock {\em Nature Physics}, (11):124--130, 2015.

\bibitem{2016_Alessio_AiP}
L.~D'Alessio, Y.~Kafri, A.~Polkovnikov, and M.~Rigol.
\newblock From quantum chaos and eigenstate thermalization to statistical
  mechanics and thermodynamics.
\newblock {\em Advances in Physics}, 65(3):239--362, 2016.

\bibitem{2014_Rigol_PRX}
L.~D'Alessio and M.~Rigol.
\newblock Long-time behavior of isolated periodically driven interacting
  lattice systems.
\newblock {\em Phys. Rev. X}, 4:041048, Dec 2014.

\bibitem{Kim2014}
Hyungwon Kim, Tatsuhiko~N. Ikeda, and David~A. Huse.
\newblock Testing whether all eigenstates obey the eigenstate thermalization
  hypothesis.
\newblock {\em Phys. Rev. E}, 90:052105, Nov 2014.

\bibitem{Lazarides2014}
Achilleas Lazarides, Arnab Das, and Roderich Moessner.
\newblock Equilibrium states of generic quantum systems subject to periodic
  driving.
\newblock {\em Phys. Rev. E}, 90:012110, Jul 2014.

\bibitem{Mori2016}
Takashi Mori, Tomotaka Kuwahara, and Keiji Saito.
\newblock Rigorous bound on energy absorption and generic relaxation in
  periodically driven quantum systems.
\newblock {\em Phys. Rev. Lett.}, 116:120401, Mar 2016.

\bibitem{Note1}
Accurate descriptions of generic undriven thermalized systems using random
  matrix theory is in general only possible over small energy windows far from
  the energy-spectrum edges. If one analyses the entire energy spectrum, the
  local structure typically encountered in static Hamiltonians emerges and
  random matrix theory fails to capture it. This is not the case for driven
  thermalized systems which accurately applies to the entire $\protect \hat
  U_F$ spectrum.

\bibitem{2020_jens}
Jens Eisert, Dominik Hangleiter, Nathan Walk, Ingo Roth, Damian Markham, Rhea
  Parekh, Ulysse Chabaud, and Elham Kashefi.
\newblock Quantum certification and benchmarking.
\newblock {\em Nature Reviews Physics}, 2(7):382--390, 2020.

\bibitem{Hangleiter2018anticoncentration}
Dominik Hangleiter, Juan Bermejo-Vega, Martin Schwarz, and Jens Eisert.
\newblock Anticoncentration theorems for schemes showing a quantum speedup.
\newblock {\em {Quantum}}, 2:65, May 2018.

\bibitem{doi:10.1137/0214060}
Larry. Stockmeyer.
\newblock On approximation algorithms for \# p.
\newblock {\em SIAM Journal on Computing}, 14(4):849--861, 1985.

\bibitem{Note2}
Even though there are recent breakthroughs worst-to-average case reduction for
  hardness of computing output probabilities of quantum circuits. So far none
  of the approaches matches the realistic error requirement to rule out a
  classical sampler from the Stockmeyer argument \cite {bouland2019complexity,
  movassagh_cayley_2019}.

\bibitem{goldberg_guo_2017}
Leslie~Ann Goldberg and Heng Guo.
\newblock The complexity of approximating complex-valued ising and tutte
  partition functions.
\newblock {\em computational complexity}, 26(4):765–833, 2017.

\bibitem{bremner_average_2016}
Michael~J. Bremner, Ashley Montanaro, and Dan~J. Shepherd.
\newblock Average-{Case} {Complexity} {Versus} {Approximate} {Simulation} of
  {Commuting} {Quantum} {Computations}.
\newblock {\em Phys. Rev. Lett.}, 117(8):080501, August 2016.
\newblock Publisher: American Physical Society.

\bibitem{2010_haake}
Fritz Haake.
\newblock {\em Quantum Signatures of Chaos}.
\newblock Springer International Publishing, US, 2010.

\bibitem{2002_simon}
Marvin~K. Simon.
\newblock {\em Probability Distributions Involving Gaussian Random Variables}.
\newblock Springer, Boston, MA, 2002.

\bibitem{p2}
Supanut Thanasilp, Jirawat Tangpanitanon, Marc-Antoine Lemonde, Ninnat
  Dangniam, and Dimitris~G. Angelakis.
\newblock Quantum supremacy and quantum phase transitions.
\newblock {\em arXiv.2012.06459: Quantum Physics}, 2020.

\bibitem{doi:10.1080/14789940801912366}
F.~Verstraete, V.~Murg, and J.I. Cirac.
\newblock Matrix product states, projected entangled pair states, and
  variational renormalization group methods for quantum spin systems.
\newblock {\em Advances in Physics}, 57(2):143--224, 2008.

\bibitem{bouland2019complexity}
Adam Bouland, Bill Fefferman, Chinmay Nirkhe, and Umesh Vazirani.
\newblock On the complexity and verification of quantum random circuit
  sampling.
\newblock {\em Nature Physics}, 15(2):159--163, 2019.

\bibitem{movassagh_cayley_2019}
Ramis Movassagh.
\newblock Cayley path and quantum computational supremacy: {A} proof of
  average-case \${\textbackslash}\#{P}-\$hardness of {Random} {Circuit}
  {Sampling} with quantified robustness.
\newblock {\em arXiv:1909.06210 [cond-mat, physics:hep-th, physics:math-ph,
  physics:quant-ph]}, October 2019.
\newblock arXiv: 1909.06210.

\bibitem{Note3}
$f(n)=o(g(n))$ means that $f(n)/g(n) \to 0$ when $n\to \infty $.

\bibitem{beni}
Daniel~A. Roberts and Beni Yoshida.
\newblock Chaos and complexity by design.
\newblock {\em Journal of High Energy Physics}, 2017(4):121, 2017.

\bibitem{Harrow2009}
Aram~W. Harrow and Richard~A. Low.
\newblock Random quantum circuits are approximate 2-designs.
\newblock {\em Communications in Mathematical Physics}, 291(1):257--302, Oct
  2009.

\bibitem{nicholas_2018}
Nicholas~R. Jones.
\newblock Chaos and randomness in strongly-interacting quantum systems.
\newblock {\em Dissertation (Ph.D.), California Institute of Technology.},
  2018.

\end{thebibliography}
\pagebreak
\begin{widetext}

\appendix
\section{Proof of the main theorem}\label{app:average-hardness}
In this section, we provide a detailed proof of the main theorem of the main text, which reads:
\bigskip
\begin{m_thm}\label{thm:main}
	Assuming conjecture 1 and 2, the ability to classically sample from $p_M(\z)$ up to an additive error $\beta=1/(8e)$ for all unitary matrices in $\{\hat{U}_F\}$ implies the collapse of the polynomial hierarchy to the third level.
\end{m_thm}
\bigskip

The proof relies on the theorems 1 and 2 and conjectures 1 and 2 presented in the main text. 

\bigskip
\begin{thm1}\label{thm:sharp_P} Let $\mathcal{Y}$ be a set of output probabilities $\tilde{p}_M(\bold{z})=|\langle \bold{z}|\hat{U}^M_{\rm COE}|\bold{z}_0\rangle|^2$ obtained from all possible COE matrices $\{\hat{U}_{\rm COE}\}$ and all possible output strings $\z\in\mathcal{Z}$. Approximating $\tilde{p}_M(\bold{z})$ in $\mathcal{Y}$ up to multiplicative error is $\shap$ hard in the worst case.
\end{thm1}
\begin{thm2} The distribution of $\tilde{p}_M(\bold{z})$ in $\mathcal{Y}$ anticoncentrates with $\delta =1$ and $\gamma=1/e$, where $e$ is the base of the natural logarithm.
\end{thm2}
\begin{conjecture_app}[Average-case hardness]\label{conj:average-hard} For any 1/(2e) fraction of $\mathcal{Y}$
approximating $\tilde{p}_M(\z)$ up to multiplicative error with $\alpha= 1/4+o(1)$, where $o(\cdot)$ is little-o notation\footnote{$f(n)=o(g(n))$ means that $f(n)/g(n) \to 0$ when $n\to\infty$.}, is as hard as the hardest instance.
\end{conjecture_app}

\begin{conjecture_app}[Computational ETH]\label{conj:ETH}
The experimentally accessible set $\{\hat{U}_F\}$ is approximately Haar random over the ensemble  $\{\hat{U}_{\rm COE}\}$ in the sense that
\begin{enumerate}
\item The distribution of $p_M({\bf z})$ over $\hat{U}_F$ is the same as that of $\tilde{p}_M(\bf z)$ over $\hat{U}_{\rm COE}$. 
\item Average instances in $\{\hat{U}_F\}$ are as hard as average instances in $\{\hat{U}_{\rm COE}\}$.
\end{enumerate}
\end{conjecture_app}

\bigskip

Let us begin by considering a classical probabilistic computer with an NP oracle, also called a $\mathrm{BPP^{NP}}$ machine. This is a theoretical object that can solve problems in the third level of the polynomial hierarchy. The Stockmeyer theorem states that a $\mathrm{BPP^{NP}}$ machine with an access to a classical sampler $\mathcal{C}$, as defined in the main text, 
can efficiently output an approximation $\tilde{q}(\z)$ of $q(\z)$ such that
\begin{align}
	|q(\z)-\tilde{q}(\z)| \le \frac{q(\z)}{\poly{L}}.
\end{align}
We emphasise that the $\mathrm{BPP^{NP}}$ machine grants us the ability to perform the approximating task, in contrast to the machine $\mathcal{C}$ that can only sample strings from a given distribution. To see how the $\mathrm{BPP^{NP}}$ machine can output a multiplicative approximation of $p_M(\z)$ for most of $\z\in\mathcal{Z}$, let us consider
\begin{align}\label{eq:long}
	|p_M(\z)&-\tilde{q}(\z)| \nonumber \\ &\le |p_M(\z)-q(\z)| + |q(\z)-\tilde{q}(\z)| \nonumber \\
		&\le |p_M(\z)-q(\z)| + \frac{q(\z)}{\poly{L}} \nonumber \\
		&\le |p_M(\z)-q(\z)| + \frac{|p_M(\z) - q(\z)| + p_M(\z)}{\poly{L}} \nonumber \\
		&= \frac{p_M(\z)}{\poly{L}} + |p_M(\z)-q(\z)| \left(1 + \frac{1}{\poly{L}}\right). 
\end{align}
The first and the third lines are obtained using the triangular inequality. To get multiplicative approximation of $p_M(\z)$ using $\tilde{q}(\z)$, we need the term $|p_M(\z)-q(\z)|$ to be small. Given the additive error defined in Eq. (3) in the main text, this is indeed the case for a large portion of $\{\z\}\in\mathcal{Z}$. Since the left hand side of Eq. (3) in the main text involves summing over an exponentially large number of terms but the total error is bounded by a constant $\beta$, most of the terms in the sum must be exponentially small. This statement can be made precise using Markov's inequality.

\bigskip
\begin{fact}[Markov's inequality] If $X$ is a non-negative random variable and $a>0$, then the probability that $X$ is at least $a$ is
\begin{equation}
{\rm Pr}(X\ge a) \le \frac{\mathbb{E}(X)}{a},	
\end{equation}
where $\mathbb{E}(X)$ is the expectation value of $X$.
\end{fact}
By setting $X=|p_M(\z)-q(\z)|$, we get
\begin{align}
\Pr_{\z} \left( |p_M(\z)-q(\z)| \ge a \right) \le \frac{\mathbb{E}_{\z}(|p_M(\z)-q(\z)|)}{a},
\end{align}
Here, the distribution and the expectation value are computed over $\z\in\mathcal{Z}$. Note that $\mathbb{E}_{\z}(|p_M(\z)-q(\z)|)\leq \beta/N$ is given by the additive error defined in Eq. (3) in the main text. By setting $a=\beta/N\zeta$ for some small $\zeta>0$, we get
\begin{align}
\Pr_{\z} \left( |p_M(\z)-q(\z)| \ge \frac{\beta}{N\zeta} \right)\le \zeta 
\end{align}
or equivalently
\begin{align}
\Pr_{\z} \left( |p_M(\z)-q(\z)| < \frac{\beta}{N\zeta} \right) > 1- \zeta. 
\end{align}
By substituting $|p_M(\z)-q(\z)|$ from Eq. (\ref{eq:long}), we get
\begin{align}\label{eq:stock1}
	\Pr_{\z}\left(|p_M(\z)-\tilde{q}(\z)| < \frac{p_M(\z)}{\poly{L}} + \frac{\beta}{N\zeta} \left(1 + \frac{1}{\poly{L}}\right) \right) > 1-\zeta. 
\end{align}
Theorem 2 in the main text (the anti-concentration condition) together with conjecture 2 imply that $\{p_M({\bf z})\}$ follows the Porter-Thomas distribution, specially that $1/N<p_M(\z)$ for at least $1/e$ fraction of the unitary matrices in $\{\hat{U}_F\}$. Hence, we can rewrite Eq. (\ref{eq:stock1}) as
\begin{align}\label{eq:stock2}
	&\Pr_{\mathcal{Y}}\left\{|p_M(\z)-\tilde{q}(\z)|< 
	p_M(\z)\left[
	 \frac{1}{\poly{L}} + \frac{\beta}{\zeta} \left(1 + \frac{1}{\poly{L}}\right) \right]\right\}  >1/e-\zeta. 
\end{align}
Here, the distribution is over all $\z\in\mathcal{Z}$ and all unitary matrices in $\{\hat{U}_F\}$. To understand the right hand side of the equation, let $P\cap Q$ be the intersection between the set $P$ of probabilities that anticoncentrate and the set $Q$ of probabilities that satisfy the Markov's inequality. Since $\Pr(P\cap Q) = \Pr(P) + \Pr(Q) - \Pr(P\cup Q) \ge \Pr(P) + \Pr(Q) -1$, $\Pr(P)=1/e$ and $\Pr(Q)=1-\zeta$ , it follows that  $\Pr(P\cap Q)$ is no less than $1/e + 1 - \zeta -1 =1/e-\zeta$.

Following \cite{bremner_average_2016, 2018_hartmut_natphy}, we further set $\beta = 1/(8e)$ and $\zeta = 1/(2e)$, so that
\begin{align}\label{eq:stock3}
	\Pr_{\hat{U}_F,\z}\left\{|p_M(\z)-\tilde{q}(\z)| <
	\left(\frac{1}{4} + o(1)\right) p_M(\z)\right\} > \frac{1}{2e},
\end{align}
giving an approximation up to multiplicative error $1/4+o(1)$ for at least $1/(2e)$ instances of the set of experimentally realizable unitary matrices $\{\hat{U}_F\}$.
If according to the conjecture \ref{conj:average-hard} and conjecture 2 in the main text, multiplicatively estimating $1/(2e)$ fraction of the output probabilities from $\{\hat{U}_F\}$ is \#P-hard,
then the Polynomial Hierarchy collapses. This concludes the proof of the main theorem in the main text.




\section{Mapping of approximating output distribution of COE dynamics onto estimating partition function of complex Ising models}\label{app:mapping}

In this section, we provide evidence to support the conjecture 1 in the main text, showing how hardness instances could appear on average. To do this, we map the task of approximating an output distributions of COE dynamics onto calculating the partition function of a classical Ising model which is widely believed to be \#P-hard on average for multiplicative approximation \cite{2018_eisert_prx,bremner_average_2016}. The section is divided into two parts. In the first part, we explain the overall concept and physical intuition of this procedure. In the second part, mathematical details are provided.

\subsection{Physical perspective of the mapping procedure}

The mapping protocol consists of two intermediate procedures. First, we map the COE unitary evolution on \textit{universal} random quantum circuits and, second, we derive a complex Ising model from those circuits following Ref. \cite{2018_hartmut_natphy}. 

Let us begin by expressing an unitary evolution of COE as $\hat{U}_{\rm COE} = \hat{U}^T_{\rm CUE}\hat{U}_{\rm CUE}$ where $\hat{U}_{\rm CUE}$ is a random unitary drawn from the Circular Unitary Ensemble (CUE) i.e. Haar ensemble \cite{2010_haake}. We then further decompose $\hat{U}_{\rm CUE}$ into a set of universal quantum gates \cite{2018_hartmut_natphy}. Following Ref. \cite{2018_hartmut_natphy}, we choose random quantum circuits consisting of $n+1$ layers of gates and $\log_2 N $ qubits, as shown in Fig. \ref{fig1}(a). The first layer consists of Hadamard gates applied to all qubits. The following layers consist of randomly chosen single-qubit gates from the set $\{\sqrt{{X}},\sqrt{{Y}},{T}\}$ and two-qubit controlled-Z (CZ) gates. Here, $\sqrt{{X}}$ ($\sqrt{{Y}}$) represents a $\pi/2$ rotation around the ${X}$ (${Y}$) axis of the Bloch sphere and $\hat{T}$ is a non-Clifford gate representing a diagonal matrix $\{1,e^{i\pi/4}\}$. Such circuits have been shown to be approximately $t$-design \cite{beni} for an arbitrary large $t$ when $n\to\infty$, which implies the CUE evolution \cite{Harrow2009}. The operator $\hat{U}^T_{\rm CUE}$ can be implemented by reversing the order of the gates in $\hat{U}_{\rm CUE}$ and replacing $\sqrt{ Y}$ with $\sqrt{Y}^T$. We emphasize that decomposing the COE evolution into the random circuits is only done theoretically with an aim to show the average case hardness. In the real experiments, this COE dynamics is realized by the driven many-body systems.

\begin{figure*}
\includegraphics[width=17.5cm,height=13.5cm]{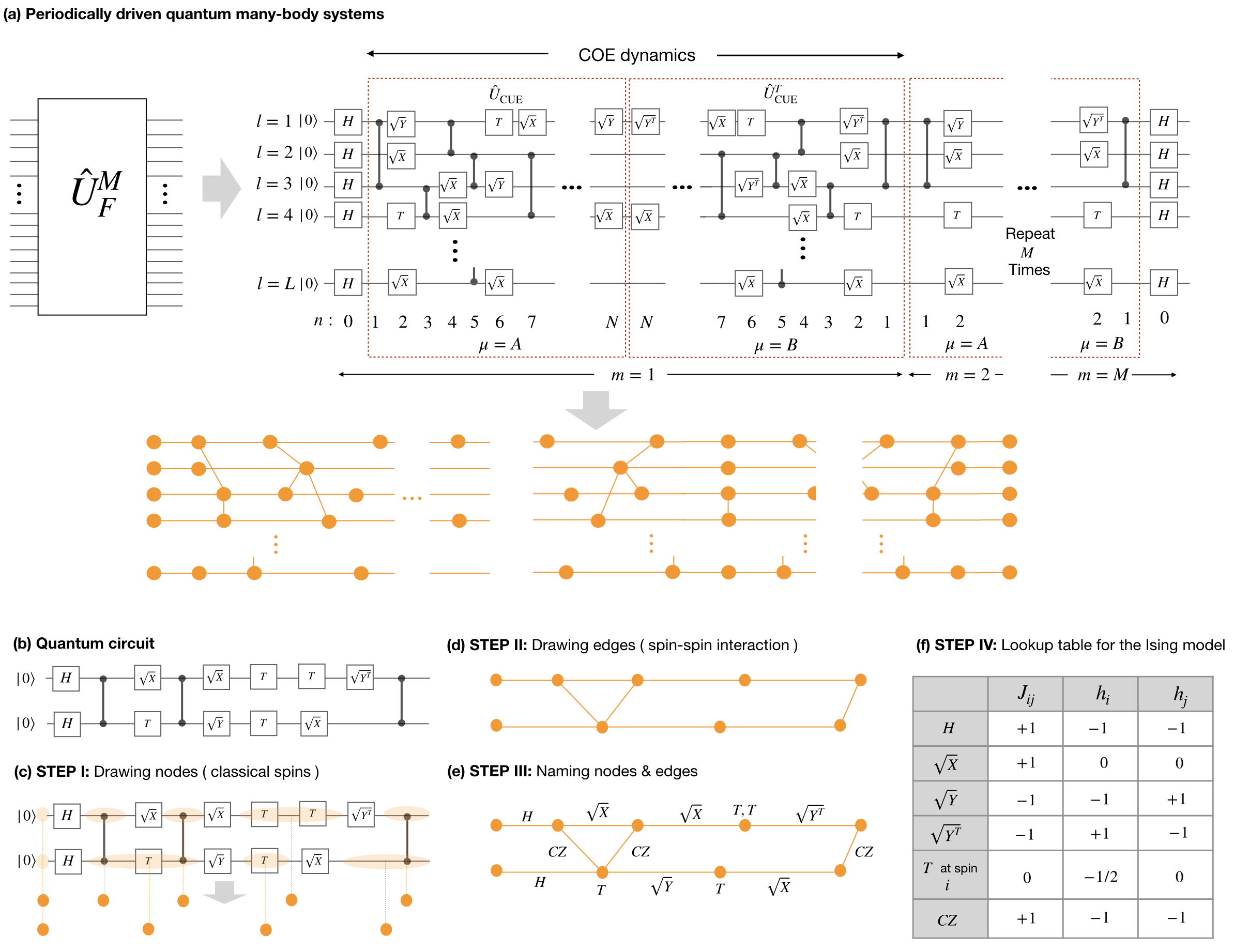}
\caption{\textbf{Mapping driven many-body dynamics to the partition function of complex Ising lattices:} (a) An example of a random circuit that generates COE dynamics and its conversion to the Ising model. (b) An example of a simple random quantum circuit, illustrating the mapping to the classical Ising model. STEP I to STEP III in the diagrammatic procedure are shown in (b)-(d), respectively. (e) Lookup table for the contribution of each gate to the local fields $h_i$, $h_j$ and the interaction $J_{ij}$ in the Ising lattice.}
\label{fig1}
\end{figure*}
 
 The mathematical procedure for the mapping from random quantum circuits to classical complex Ising models is discussed in details in the next part. Specifically, $p_M(\bold{z})$ from the circuit $(\hat{U}^T_{\rm CUE}\hat{U}_{\rm CUE})^M$, as depicted in Fig. \ref{fig1}(a), can be calculated from the partition function,
\begin{equation}
\langle \bold{z} |\hat{U}^M_{\rm COE}|\z_0\rangle = \sum_{\bold{s}\in \mathcal{S}} A(\bold{s})\exp \left[\frac{i\pi}{4} \left(\sum_{i} h_i s_i+\sum_{\langle i,j\rangle}J_{ij}s_is_j\right)\right].
\label{eq:ising_partition}
\end{equation}
Here, $A(\bold{s})$ is the degeneracy number associated with a classical spin configuration $\bold{s}$ in the lattice $\mathcal{S}$, $s_i=\pm 1$, $h_i$ represents a on-site field on site $i$ and $J_{ij}$ represents the coupling between the classical spins on site $i$ and $j$. Since the output probability can also be interpreted as the path integral in Eq.~(\ref{eq:ising_partition}) in the main text, the intuition behind the mapping is that the sum over all possible paths is translated into the sum over all possible classical spin configurations, where the phase accumulated in each path is given by the energy of the complex Ising lattice $\mathcal{S}$. To gain intuitive understanding of this standard mapping, we provide a diagrammatic approach to visualize the lattice $\mathcal{S}$ and extract the field parameters $\{h_i\}$, $\{J_{ij}\}$. To begin with, we use the random circuit in Fig. \ref{fig1}(b) as a demonstration. The mathematical descriptions behind each steps are discussed in the next part.
\begin{itemize}
	\item \underline{STEP I} - For each qubit, draw a circle between every consecutive non-diagonal gates, see Fig. \ref{fig1}(c). Each circle or `node' represents one classical spin. 
	\item \underline{STEP II} - For each qubit, draw a horizontal line between every consecutive nodes $i$,$j$, see Fig. \ref{fig1}(d). These lines or `edges' represent interaction $J_{ij}$ between two neighboring spins in the same row. In addition, draw a line between every two nodes that are connected by $CZ$ gates. These lines represent the interaction $J_{ij}$ between spins in different rows. 
	\item \underline{STEP III} - Labeling each nodes and edges with the corresponding gates, see Fig. \ref{fig1}(e).  
	\item \underline{STEP IV} - Use the lookup table in Fig. \ref{fig1}(f) to specify $h_i$ and $J_{ij}$ introduced by each gate. For example, the $\sqrt{Y}$ gate that acts between nodes $i$ and $j$ adds $-1$ to $J_{ij}$, $-1$ to $h_i$ and $+1$ to $h_j$. We use the convention that the leftmost index represents the leftmost node. Also, the two T-gates that are enclosed by the node $i$ will add $0.5+0.5=+1$ to the local field $h_i$.
	\item \underline{STEP V} - Finally, spins at the leftmost side of the lattice are fixed at $+1$, corresponding to the initial state $|\bold{0}\rangle$. Similarly, spins at the rightmost side of the lattice are fixed according to the readout state $\ket{\bold{z}}$.
\end{itemize}

Following the above recipe, we provide the exact form of the parameters in the Ising model for the COE dynamics in the next part, showing that the field parameters $\{h_i\}$ and $\{J_{ij}\}$ are quasi-random numbers with no apparent structure. Specifically, neither the phase $\pi \sum_i h_i s_i /4$ nor the phase $\pi \sum_{\langle i,j\rangle} J_{ij} s_i s_j/4$ is restricted to the values $0,\pi/2,\pi,3\pi/2$ (mod $2\pi$) for each spin configurations $\bold{s}$. Without such stringent restrictions, approximating the partition function up to multiplicative error is known to be $\shap$-hard in the worst case \cite[Theorem 1.9]{goldberg_guo_2017}. This motivates a widely used conjecture in quantum supremacy proposals that such task is also hard on average \cite{2018_eisert_prx,bremner_average_2016}. 


We emphasize here the major differences between random quantum circuits as proposed in Ref. \cite{2018_hartmut_natphy} and our systems. Firstly, our systems are analog with no physical quantum gates involved. The decomposition to quantum gates is only done mathematically. Secondly, our system has discrete time-reversal symmetry, while such symmetry is absent in random quantum circuits. Consequently, the COE in our system is achieved from the Floquet operator $\hat{U}_F$, while the CUE in random quantum circuits are achieved from the entire unitary evolution. In addition, $\hat{U}_F^M$ in our system does not have the $t$-design property due to the COE \cite[pp.117-119]{nicholas_2018}. However, as shown above, the hardness arguments for the random quantum circuits can be naturally applied to our case.

\subsection{Mathematical details of the mapping procedure}

In this section, we prove Eq.~(\ref{eq:ising_partition}) by providing justifications of the diagrammatic recipes to map the the evolution $\hat U_{\rm CUE}$ on a Ising spin model with complex fields. Again, the quantum gates of interest consist of both diagonal gates $\{T,CZ\}$ and non-diagonal gates $\{\sqrt{X}, \sqrt{Y}, \sqrt{Y}^T,H\}$. For simplicity, we start with one- and two- qubit examples before generalizing to the COE dynamics. The mathematical procedure here is adapted from Ref. \cite{2018_hartmut_natphy}.

\subsubsection{One-qubit example}

Let us consider a one-qubit circuit and $N+1$ gates randomly chosen from the set $\{\sqrt{X},\sqrt{Y},\sqrt{Y}^T,T\}$. The zeroth gate is fixed to be a Hadamard gate. The output probability is $p(z)=|\langle z |\hat{U}|0\rangle|^2$, where $\hat{U}=\prod_{n=0}^N \hat{U}^{(n)}$ is the total unitary matrix, $\hat{U}^{(n)}$ is the $n^{\rm th}$ gate and $z\in\{0, 1\}$ is the readout bit. Below, we outline the mathematical steps underlying the diagrammatic approach followed by detailed explanations for each step:
\begin{align}
		p(z) & = \left|\bra{z} \prod_{n=0}^{N} \hat{U}^{(n)} \ket{0}\right|^2 \nonumber\\
			 & = \left|\sum_{\underline{z}\in\{0,1\}^{N}} \prod_{n=0}^{N} \bra{z_{n}} \hat{U}^{(n)} \ket{z_{n-1}} \right|^2\nonumber\\
			 & = \left|\sum_{\underline{z}\in\{0,1\}^{N}} \prod_{n=0}^{N} A(z_{n},z_{n-1}) \exp\left[\frac{i\pi}{4}\Phi(z_{n},z_{n-1})\right]\right|^2 \nonumber\\
			& = \left|\sum_{\underline{z}\in\{0,1\}^{N+2}} A({\underline{z}}) \exp\left[\frac{i\pi}{4} \sum_{n=0}^{N}\Phi(z_{n},z_{n-1})\right]\right|^2.
\label{eqA1}
\end{align}
In the second line, we insert an identity $\hat{I}_n=\sum_{z_n\in\{0,1\}} \ket{z_n} \bra{z_n}$ between $\hat{U}^{(n+1)}$ and $\hat{U}^{(n)}$ for every $n\in\{0,..,N-1\}$. As a result, this line can be interpreted as the Feynman's path integral where each individual path or `world-line' is characterized by a sequence of  basis variables $\underline{z} = (z_{-1},z_0,...,z_{N})$. The initial and the end points for every path are $\ket{z_{-1}} = \ket{0}$ and $\ket{z_N} = \ket{z}$, respectively. In the third line, we decompose $\bra{z_{n}} \hat{U}^{(n)} \ket{z_{n-1}}$ into the amplitude $A(z_{n},z_{n-1})$ and phase $\Phi(z_{n},z_{n-1})$. In the fourth line, we introduce $A({\underline{z}})=\prod_{n=0}^{N} A(z_{n},z_{n-1})$. The equation now takes the form of the partition of a classical Ising model with complex energies. Here, $\underline{z}$ can be interpreted as a classical spin configuration, $A(\underline{z})$ as the degeneracy number and $i\frac{\pi}{4}\Phi(z_n,z_{n-1})$ as a complex energy associated with spin-spin interaction. 

Further simplifications are possible by noticing that, the diagonal gates in the circuits allow the reduction of the number of classical spins. Specifically, if a $T$ gate is applied to $\ket{z_{n-1}}$, it follows that $z_n=z_{n-1}$. Hence, the variables $z_{n-1}$ and $z_{n}$ can be represented by a single classical spin state. The two variables $z_{n-1}, z_{n}$  become independent only when a non-diagonal gate is applied. Therefore, we can group all variables $\{z_n\}$ between two non-diagonal gates as one classical spin. This procedure leads to the directives presented as the the STEP I of the procedure in the previous section. Formally, for $N_{\rm spin}+1$ non-diagonal gates in the circuit (including the first Hadamard gate) $\underline{z}$ can be characterized by a classical spin configuration $\underline{s} = (s_{-1},s_0,...,s_k,...,s_{N_{\rm spin}})$ where $s_{k}=1-2z_k\in\{\pm 1\}$ is a spin representing the basis variable immediately after the $k^{th}$ non-diagonal gate, i.e.

\begin{align}
		p(z) & = \left|\sum_{\underline{s} \in \{\pm 1\}^{N_{\rm spin}+1}}  A({\underline{s}})\exp{\left[\frac{i\pi}{4}\sum_{k=0}^{N_{\rm spin}}\Phi(s_{k},s_{k-1})\right]}\right|^2 \\
			& = \left|Z_{\rm Ising}\right|^2 \\
\end{align}

Lastly, we need to specify $A(\underline{s})$ and $\Phi(s_{k},s_{k-1})$ in term of the local fields $h_{k-1}$, $h_k$, the interaction $J_{k-1,k}$, and spin configurations $s_{k-1},s_{k}$. This is done by first considering the gates in their matrix form, i.e.

\begin{align}
		\sqrt{X} & = \frac{1}{\sqrt{2}}\begin{pmatrix}
			e^{\frac{i\pi}{2}} & 1 \\
			1  & e^{\frac{i\pi}{2}} 
		\end{pmatrix} = \frac{1}{\sqrt{2}}\left[ e^{\frac{i\pi}{4}(1+s_ks_{k-1})}\right]_{s_k,s_{k-1}}, \\
		\sqrt{Y} & = \frac{1}{\sqrt{2}}\begin{pmatrix}
			1 & -1 \\
			1  & 1 
		\end{pmatrix} = \frac{1}{\sqrt{2}}\left[ e^{\frac{i\pi}{4}(1-s_{k-1})(1+s_{k})}\right]_{s_k,s_{k-1}}, \\
		\sqrt{Y}^T & = \frac{1}{\sqrt{2}}\begin{pmatrix}
			1 & 1 \\
			-1  & 1 
		\end{pmatrix} = \frac{1}{\sqrt{2}}\left[ e^{\frac{i\pi}{4}(1+s_{k-1})(1-s_{k})}\right]_{s_k,s_{k-1}}, \\
		H & = \frac{1}{\sqrt{2}}\begin{pmatrix}
			1 & 1 \\
			1 & -1 
		\end{pmatrix} = \frac{1}{\sqrt{2}}\left[ e^{\frac{i\pi}{4}(1-s_{k-1})(1-s_{k})}\right]_{s_k,s_{k-1}}, \\
		T & = \begin{pmatrix}
			1 & 0 \\
			0  & e^{\frac{i\pi}{4}} 
		\end{pmatrix} = {\rm Diag}\left[ e^{\frac{i\pi}{4}(\frac{1-s_{k}}{2})}\right]_{s_k}
\end{align}

Notice that all non-diagonal gates contribute to the same amplitude $A(s_k,s_{k-1})=1/\sqrt{2}$, leading to $A(\underline{s})=2^{-(N_{\rm spin}+1)/2}$. Hence, we can extract the contribution of each gate to $\Phi(s_k,s_{k-1})$ as 
\begin{align}
		\Phi_{\sqrt{X}}(s_k,s_{k-1}) & = 1+s_{k-1}s_{k},\\
		\Phi_{\sqrt{Y}}(s_k,s_{k-1}) & = (1-s_{k-1})(1+s_{k})
		\\ &= 1-s_{k-1}+s_{k}-s_{k-1}s_k,\\
		\Phi_{\sqrt{Y}^T}(s_k,s_{k-1}) & = (1+s_{k-1})(1-s_{k})
		\\ &= 1+s_{k-1}-s_k-s_{k-1}s_k  ,\\
		\Phi_T(s_{k}) & = \frac{1-s_{k}}{2}.
\end{align}

The under-script indicates which gate is contributing to the phase. The corresponding $h_i$, $h_j$ and $J_{ij}$ are depicted in the lookup table in Fig. \ref{fig1}(f), where $i=k-1$ and $j=k$. The global phase that does not depend on $\underline{s}$ is ignored as it does not contribute to $p(z)$. 


\subsubsection{Two-qubit example}
Now we consider a two-qubit random circuits to demonstrate the action of the $CZ$ gates. We introduce a new index $l \in\{1,2\}$ to label each qubit, which is placed on a given horizontal line (row). Since the $CZ$ gate is diagonal, its presence does not alter the number of spins in each row. However, the gate introduces interaction between spins in different rows. This can be seen from its explicit form, i.e.

\begin{align}
		CZ & = \begin{pmatrix}
			1 & 0 & 0& 0 \\
			0 & 1 & 0& 0 \\
			0 & 0 & 1& 0 \\
			0 & 0 & 0& -1
		\end{pmatrix} = {\rm Diag}\left[ e^{\frac{i\pi}{4}(1-s_{1,k})(1 - s_{2,k'})}\right]_{s_{1,k},s_{2,k'}},
\end{align}

where $s_{1,k}$ ($s_{2,k'}$) is the state of the $k^{\rm th}$ ($k'^{\rm th}$) spin at the first (second) row. It follows that

\begin{align}
		\Phi^{CZ}_{s_{1,k},s_{2,k'}} &= (1-s_{1,k})(1 - s_{2,k'}) \\
		&= 1 - s_{1,k} - s_{2,k'} + s_{1,k}s_{2,k'}.
\end{align}

The corresponding $h_i$, $h_j$, and $J_{ij}$ are depicted in Fig. \ref{fig1}(f) where $i=(1,k)$ and $j=(2,k')$. We have now derived all necessary ingredients to map a random quantum circuit to a classical Ising model.


\subsubsection{Full COE dynamics}

Since the COE dynamics can be expressed in terms of a quasi-random quantum circuit, we can straightforwardly apply the above procedure to find the corresponding Ising model. The complexity here solely arises from the number of indices required to specify the positions of all the gates in the circuit. To deal with this, we introduce the following indices
\begin{itemize}
\item[--] an index $l\in\{1,...,L\}$ to indicate which qubit / row.
\item[--] an index $m\in\{1,...,M\}$ to indicate which period.
\item[--] an index $\mu\in\{A,B\}$ to indicate which part of the period. $A$ and $B$ refer to the $\hat{U}_{\rm CUE}$ part and the $\hat{U}_{\rm CUE}^T$ part, respectively
\item[--] an index $k\in\{0,1,...,N_{\rm spin}(l)\}$ to indicate the spin position for a given $m$ and $\mu$. Here, $N_{\rm spin}(l)$ is the total number of spins at the $l^{\rm th}$ row. Note that due to the symmetric structure of $\hat{U}_{\rm CUE}$ and $\hat{U}_{\rm CUE}^T$, we run the index $k$ backward for the transpose part, i.e. $k=0$ refers to the last layer.
\item[--] an index $\nu_{l,k}$ so that $\nu_{l,k}=1$ if the $k^{\rm th}$ non-diagonal gate acting on the qubit $l$ is $\sqrt{X}$ otherwise $\nu_{l,k}=0$.
\end{itemize}
With these indices, the partition function of the circuit, as shown in Fig. \ref{fig1}(a), can be written as
\begin{equation}
	\bra{\bold{z}} \psi \rangle = 2^{-\frac{G}{2}}\sum_{\underline{s}\in \mathcal{S}} \exp{\left[\frac{i\pi}{4}E(\textbf{s})\right]},
\end{equation}
with
\begin{align}
		E(\textbf{s}) = & E(\bold{z}) + \sum_{m=1}^{M}\sum_{\mu=A}^{B}\sum_{l=1}^{L}\sum_{k=0}^{N_{\rm spin}(l)} h_{lk} s^{\mu,m}_{l,k} \\ 
		&+\sum_{m=1}^{M}\sum_{\mu=A}^{B}\sum_{l=1}^{L}\sum_{k=1}^{N_{\rm spin}(l)} (2\nu_{l,k}-1)s^{\mu,m}_{l,k-1}s^{\mu,m}_{l,k} \nonumber \\ 
		& + \sum_{m=1}^{M}\sum_{\mu=A}^{B }\sum_{l=1}^{L}\sum_{l'=1}^{l-1}\sum_{k=1}^{N_{\rm spin}(l)}\sum_{k'=1}^{N_{\rm spin}(l')} \zeta_{(l,k)}^{(l',k')} s^{\mu,m}_{l,k} s^{\mu,m}_{l',k'} \nonumber,
\end{align}
and
\begin{align}
		h_{lk} & = \nu_{l,k+1} - \nu_{l,k} - \frac{1}{2} N_T{(l,k)} - N_{CZ}(l,k), \\
		E(\bold{z}) & = -s^{B,M}_{0,l} - s_{z_{l}} + s^{B,M}_{0,l}s_{z_{l}}.
\end{align}
Here $G$ is the total number of non-diagonal gates in the circuit. $\zeta_{(l,k)}^{(l',k')}$ represents the total number of $CZ$ gates which introduces the interaction between spins $s^{\mu,m}_{l,k}$ and $s^{\mu,m}_{l',k'}$.  $N_{CZ}(l,k)$ ($N_T{(l,k)}$) is the total number of $CZ$ ($T$) gates which introduces local fields on the spin $s^{\mu,m}_{l,k}$. $E(\bold{z})$ is the contribution from the last Hadamard layer which depends on the readout bit-string $\bold{z}$. $\{s_{z_l}\}$ are the spins corresponding to $\bold{z}$ and their configuration is fixed. In addition, there are also two extra boundary conditions (i) between part $A$ and $B$ and (ii) between the two adjacent periods $m$ and $m+1$, i.e. $s^{A,m}_{l,N_{\rm spin}(l)} = s^{B,m}_{l,N_{\rm spin}(l)}$ and $s^{A,m+1}_{l,0} = s^{B,m}_{l,0}$. 

\section{Derivation of Porter-Thomas distribution from COE dynamics.}\label{app:anti-concentration}
In this section, we provide additional mathematical details involved in the proof of theorem 2. More precisely, we show that the distribution of the output probability of COE dynamics, ${\rm Pr} (p)$, follows the Porter-Thomas distribution ${\rm Pr_{PT}}(p)=Ne^{-Np}$. First, let us consider the output probability $p_M(\bold{z})= |\bra{\bold{z}} \psi_M \rangle|^2$ with
\begin{align}
\bra{\bold{z}} \psi_M \rangle & = \bra{\bold{z}} U^M_{\rm COE} \ket{\bold{0}} \nonumber \\
& = \bra{\bold{z}} \left[ \sum_{\epsilon=0}^{N-1} e^{iME_{\epsilon}T} \ket{E_{\epsilon}} \bra{E_{\epsilon}} \right]\ket{\bold{0}}\nonumber  \\
& = \sum_{\epsilon=0}^{N-1} d_\epsilon(\bold{z})e^{i\phi_{M,\epsilon}} \nonumber \\
& = \left[\sum_{\epsilon=0}^{N-1} d_\epsilon(\bold{z}) \cos\phi_{M,\epsilon}\right] + i \left[\sum_{\epsilon=0}^{N-1} d_\epsilon(\bold{z}) \sin\phi_{M,\epsilon}\right] \nonumber \\
& = a_{\bold{z}} + i b_{\bold{z}}, 
\end{align}
where $N$ is the dimension of the Hilbert space, $d_{\epsilon}(\bold{z})=\langle \bold{z} | E_{\epsilon}\rangle \langle E_{\epsilon}|\bold{0}\rangle$, $\phi_{m,\epsilon}=ME_\epsilon T \text{ mod }2\pi$, $a_{\bold{z}}={\rm Re}\left[\bra{\bold{z}} \psi_M \rangle\right]$ and $b_{\bold{z}}={\rm Im}\left[\bra{\bold{z}} \psi_M \rangle\right]$.
\bigskip
\begin{lem} \label{theorem1} The distribution of $d_{\epsilon}(\bold{z})$ over $\forall\epsilon\in\{0,...,N-1\}$ or $\forall \bold{z}\in\{0,1\}^L$ is the Bessel function of the second kind. \end{lem}
\begin{lem} \label{theorem2} The distribution of $a_{\bold{z}}$ and $b_{\bold{z}}$ over $\forall \bold{z}\in\{0,1\}^L$ is the normal distribution with zero mean and variance equal to $1/2N$. \end{lem}
\bigskip

To prove lemma 1, we first write $d_{\epsilon}(\bold{z}) =  c_{\bold{z},\epsilon} c_{\bold{0},\epsilon}$, where $c_{\bold{z},\epsilon}=\langle \bold{z}|E_{\epsilon}\rangle$ and $c_{\bold{0},\epsilon}=\langle \bold{0}|E_{\epsilon}\rangle$. For the COE dynamics, the coefficients $c_{\bold{z},\epsilon}$ and $c_{\bold{0},\epsilon}$ are real numbers whose distribution is \cite{2010_haake}
\begin{equation}
\rm{Pr}(c)=\sqrt{\frac{2N}{\pi}}\exp\left[-\frac{Nc^2}{2}\right].
\end{equation}
As discussed in the main text, the phase $\phi_{M,\epsilon}$ becomes random as $M\gg 2\pi /E_\epsilon T$. The random sign ($\pm 1$) from $c_{z,\epsilon}$ can therefore be absorbed into the phase without changing its statistics. The distribution of $d_{\epsilon}(\bold{z})$ can be obtained using the product distribution formula
\begin{align}
\rm{Pr}(d) & = \int_0^{\infty} \rm{Pr}(c)  \rm{Pr}(\frac{d}{c}) \cdot\frac{1}{c}\cdot \rm{d} c \nonumber \\
& =  \frac{2N}{\pi} \int_0^{\infty} \exp{\left( -\frac{Nc^2}{2}\right)} \exp{\left( -\frac{Nd^2}{2c^2}\right)}\rm{d} c \nonumber \\
& = \frac{2N}{\pi}K_0(Nd),
\end{align}
where $K_0$ is the modified Bessel function of the second kind.

To prove  lemma 2, we first note that the distribution of $\cos\phi_{m,\epsilon}$ and $\sin\phi_{m,\epsilon}$ with  $\phi_{M,\epsilon}$ being uniformly distributed in the range $[0,2\pi)$ are 

\begin{align}
\rm{Pr}(\cos\phi) & = \frac{1}{\pi\sqrt{1-\cos^2\phi}}, \\
\rm{Pr}(\sin\phi) & =  \frac{1}{\pi\sqrt{1-\sin^2\phi}}.
\end{align}
We then calculate the distribution of $\kappa_{\epsilon}\equiv d_{\epsilon}(\bold{z}) \cos\phi_{M,\epsilon}$ using the product distribution formula, i.e.
\begin{align}
\rm{Pr}(\kappa)  & = \int_{-1}^1 \frac{1}{\pi\sqrt{1-\cos^2\phi}}\cdot \frac{2N}{\pi} K_{0}(\frac{N\kappa}{d}) \cdot \frac{1}{\cos\phi} \rm{d} \cos\phi \nonumber \\
& = \frac{N}{\pi^2} K^2_0\left(\frac{N|\kappa|}{2}\right).
\end{align}
The mean and the variance of $\kappa_{\epsilon}$ can be calculated as

	\begin{align}
	\langle \kappa \rangle  & = \int_{-\infty}^\infty d\cos \phi\cdot \frac{N}{\pi^2} \cdot K^2_0\left(\frac{N|\kappa|}{2}\right)\cdot \rm{d} \kappa  = 0 \\
	{\rm Var}(\kappa) & = \int_{-\infty}^\infty (d\cos\phi)^2\cdot\frac{N}{\pi^2}\cdot K^2_0\left(\frac{N|\kappa|}{2}\right)\cdot \rm{d} \kappa  = \frac{1}{2N^2}.
	\end{align}

Since $a_{\bold{z}}$ is a sum of independent and identically distributed random variables, i.e.  $a_{\bold{z}}=\sum_{\epsilon=1}^{N-1} \kappa_\epsilon$, we can apply the central limit theorem for large $N$. Hence, the distribution of $a_{\bold{z}}$ is normal with the mean zero and variance ${\rm Var}(a) = N \cdot{\rm Var}(\kappa) = 1/2N$. The same applies for the distribution of $b_{\bold{z}}$.

Theorem 2 can be proven using the fact that the sum of the square of Gaussian variables follows the $\chi$-squared distribution with second degree of freedom $\rm{Pr}_{\chi^2,k=2}(p) \sim \exp\{-p/2\sigma^2\}$ \cite{2002_simon}. By specifying the variance obtained in Lemma 2 and normalization, the distribution of $p_M(\textbf{z}) = a_{\bold{z}}^2 + b_{\bold{z}}^2$ over $\forall \bold{z}\in\{0,1\}^L$ is the Porter-Thomas distribution. Since the Porter Thomas distribution anti-concentrates i.e. ${\rm Pr}_{\rm PT}\left(p > \frac{1}{N}\right) = \int_{Np=1}^{\infty} d(Np) e^{-Np} = 1/e$ , we complete the proof of the theorem 2.

\section{Undriven thermalized many-body systems\label{app:undriven_physical}}

In this section, we analyze the long-time unitary evolution for undriven systems in the thermalized phase. The results presented here highlight the key role played by the drive in generating the randomness required for the above quantum supremacy proof. In particular, we show that for typical undriven physical systems with local constraints (e.g. finite-range interactions) and conserved energy, the output distribution never coincides with the PT distribution. 

We emphasize that this is a consequence of the inability of random matrix theory to accurately describe the full spectral range of undriven thermalized many-body systems. Indeed, it has been shown that for undriven many-body systems which thermalizes (to a finite temperature), the statistics of the Hamiltonian resembles the statistics of the Gaussian orthogonal ensemble (GOE)~\cite{2016_Alessio_AiP}. However, it is implicit that an accurate match only applies over a small energy window (usually far from the edges of the spectrum). If one zooms in this small energy window, the Hamiltonian looks random, but if one consider the full spectrum, the local structure of the Hamiltonian appears and the random matrix theory fails at capturing it.

\begin{figure*}
\includegraphics[width=1\textwidth]{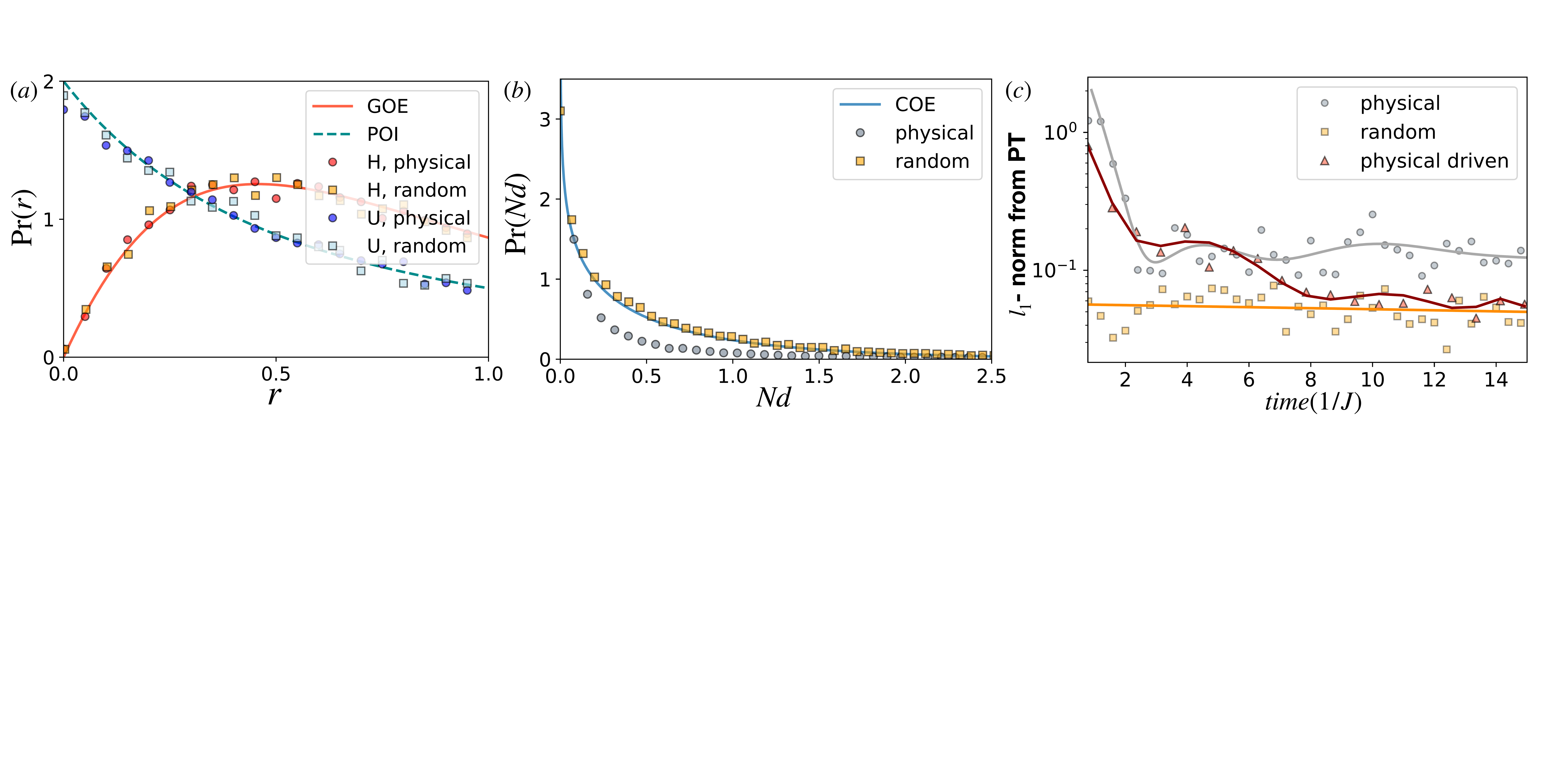}
\caption{\textbf{Undriven thermalized Ising models versus the GOE:} (a) Level-spacing statistic of an ensemble $\{ \hat H \}$ and their corresponding long-time evolution operator $\hat U$ obtained from the physical Ising system (circle) and the GOE (square). The blue dashed and the orange solid lines are theoretical predictions for the POI and the GOE distributions, respectively. (b) The eigenstate distribution $d_{\epsilon}(\bold{z})$ [see Eq. (5) of the main text] with the GOE prediction (solid line). (c) The $l_1-$norm distance between the output distribution and the PT distribution as a function of time. 
The driven case studied in the main text is presented for comparison. The parameters used are: $L = 9$, $W=1.5J, F=2.5J, \omega=8J$ (for the driven case) and $500$ disorder/instances realizations.}
\label{fig_sm}
\end{figure*}
 
To see this, we numerically simulate the undriven Ising Hamiltonian, $\hat{H}_0=\sum_{l=0}^{L-1} \mu_l \hat{Z}_l + J\sum_{l=0}^{L-2}\hat{Z}_l\hat{Z}_{l+1} + \frac{F}{2}\sum_{l=0}^{L-1}\hat{X}_l$, where $\mu_l\in\{0,W\}$ is a local disorder, $W$ is the disorder strength, $F$ is the static global magnetic field along $x$ and $J$ is the interaction strength. This Hamiltonian is in fact the average Hamiltonian of the driven Ising Hamiltonian used in the main text. In comparison, we also simulate the quantum evolution under an ensemble $\{\hat H_{\rm COE}\}$ of synthetic Hamiltonians that are uniformly drawn from the GOE (i.e. without any local constraints). 

Fig.\ref{fig_sm} (a) shows the level-spacing statistics of $\{ \hat{H}_{0} \}$ (obtained over $500$ disorder realizations), $\{ \hat{H}_{\rm COE} \}$ (obtained over $500$ random instances) and their corresponding long-time unitary operators $\hat{U} = \lim_{t\rightarrow \infty} e^{-it\hat{H}}$. We see that the level statistic of the physical Hamiltonian (and its long-time evolution) is indistinguishable from the GOE.
However, the discrepancy between the physical and synthetic (GOE) realizations becomes apparent when looking at the eigenstate statistics as shown in Fig.\ref{fig_sm} (b).
While the distribution of $d_{\epsilon}(\bold{z})$ [see Eq. (5) of the main text] from the GOE is in a good agreement with the Bessel function of the second kind, the physical system fails to meet the theoretical prediction. 
This is in contrast to the driven case as presented in the main text. 
More importantly in the context of this work, a key difference between the physical Hamiltonian and the random matrix theory prediction can be seen by comparing the distribution of the output states after some time evolution. 
In Fig.\ref{fig_sm} (c), we show that the Porter-Thomas distribution is never achieved with the physical systems while it is for the synthetic realizations as well as for the driven case studied in the main text. 
These results underline the gap between physical Hamiltonians and true random matrices and more importantly, they highlights the important role of the drive in bridging that gap.

\end{widetext}

\end{document}